\documentclass[journal,draftclsnofoot,onecolumn]{IEEEtran}
\usepackage{graphics}
\usepackage{graphicx}
\usepackage{epsfig}
\usepackage{amsmath,amssymb}
\usepackage{mathrsfs,amsfonts,fancyhdr}
\usepackage{multirow}
\usepackage{cite}
\usepackage{amsthm}

\theoremstyle{remark}

\begin{document}

\title{Achievable Rates and Outer Bound for the Half-Duplex MAC with Generalized Feedback}

\author{\authorblockN{Ahmad Abu Al Haija and Mai Vu,\\}
\authorblockA{Department of Electrical and Computer Engineering\\
McGill University\\
Montreal, QC H3A 2A7\\
Emails: ahmad.abualhaija@mail.mcgill.ca, mai.h.vu@mcgill.ca}}

\maketitle

\begin{abstract}
This paper provides comprehensive coding and outer bound for the half-duplex multiple access channel with generalized feedback (MAC-GF). Two users communicate with one destination over a discrete memoryless channel using time division. Each transmission block is divided into 3 time slots with variable durations: the destination is always in receive mode, while each user alternatively transmits and receives during the first $2$ time slots, then both cooperate to send
information during the last one. The paper proposes two decode-forward based coding schemes, analyzes their rate regions, and also derives two outer bounds with rate constraints similar to the achievable regions. Both schemes requires no block Makovity, allowing the destination to decode at the end of each block without any delay. In
the first scheme, the codewords in the third time slot are superimposed on the codewords of the first two, whereas in the second scheme, these codewords are independent. While the second scheme is simpler, the first scheme helps emphasize the importance of joint decoding over separate decoding among multiple time slots at the destination. For the Gaussian channel, the two schemes with joint decoding are equivalent, as are the two outer bounds. For
physically degraded Gaussian channels, the proposed schemes achieve the capacity. Extension to the $m$-user half-duplex MAC-GF are provided. Numerical results for the Gaussian channel shows significant rate region improvement over the classical MAC and that the outer bound becomes increasingly tight as the inter-user link quality increases.

\end{abstract}

\IEEEpeerreviewmaketitle

\section{Introduction}\label{sec:intro}

\IEEEPARstart{T}{he} growing demands of multimedia services in communication systems necessitate new technologies that meet high speed and throughput requirements. Cooperative communication offers an efficient way to increase the data rate. However, besides the high data rates demand, practical constraints including half-duplex tranceivers and short decoding delay add other challenges to new systems. Consider cooperation between two users sending information to a common destination. In \cite{fcc2005fof}, Willems et al. model this channel as a multiple access channel with generalized feedback (MAC-FG) and propose a full-duplex coding scheme that uses block Markov encoding and backward decoding. While the full-duplex scheme can be adapted to half-duplex systems, it may be inefficient. In this paper, we aim to propose new cooperative schemes directly for the half-duplex channel with shorter decoding delay.

The full-duplex MAC-GF proposed by Willems et al. in \cite{fcc2005fof} is a more practical channel than the MAC with conferencing encoders also proposed by Willems in \cite{ghasemi2007fls}. In the MAC with conferencing encoders, each encoder obtains information about the message of the other encoder through delay-free communication links between them even before starting transmission. However, in the MAC-GF, the users are cooperating during their transmission time. Willems et al. employ block Markov encoding and backward decoding to derive an achievable rate region for the MAC-GF. With backward decoding, the channel in any transmission block resembles the MAC with common message proposed by Slepian-Wolf in \cite{haykin2005crb}. In \cite{dbout}, Tandon and Ulukus derive an outer bound for the MAC-GF using the idea of dependence balance \cite{ray4} and show  that their bound is tighter than the cut-set bound. Ekrem and Ulukus \cite{lieblein1955mos} study the effects of cooperation on the secrecy of the MAC-GF.

The relay channel,  introduced by Van der Meulen in \cite{ray1}, can be seen as a special case of the MAC-GF when only one user has information to send and the other helps relay what it received to the destination. In the MAC-GF, at the end of each transmission block, each user employs partial decode-forward relaying to decode a part of the other's message and forward with its new message during the next block. Partial decode-forward is one of the relaying protocols introduced by Cover and El Gamal in \cite{ray2} along with decode-forward and compress-forward.
Kramer et al.  generalize these protocols to a relay network in \cite{kolodzy2006itm}. In \cite{ray5}, Liang and Veeravalli propose the broadcast relay channel (BRC), in which one of the two receivers assists the transmission to the other by relaying, and establish its capacity in the degraded case. Reznik et al. \cite{ray6} extend the BRC to multiple receivers and derive its achievable rate region and outer bound.

Sendonaris et al. \cite{cognitiveRT}  apply the coding scheme of the full-duplex MAC-GF into cellular networks operating over fading channels and show the advantage of cooperation in increasing both achievable rates and cellular coverage, and in reducing the outage probability. However, current technologies for cellular networks support only half-duplex communications. Hence, to make the application possible, the full-duplex scheme can be adapted to half-duplex systems using standard frequency division for channels between the two users. However, this adaptation requires extra bandwidth and thus may not be the most efficient. As a result, more attention to half-duplex schemes has been seen recently.

For example, Laneman et al. analyze the performance of half-duplex cooperative schemes in terms of  outage capacity in  \cite{sagias2004pad}, and  Vishwanath et al. derive outer bounds for the capacity of the half-duplex relay channel in \cite{simon2005dco}.  Peng and Rajan study capacity bounds for the Gaussian interference channel with transmitter or receiver cooperation in \cite{nakagami1960mdg}, and Wu et al. derive the sum capacity  for the symmetric interference channel with transmitter cooperation in \cite{turin1972smu}.  Kim et al. study the half-duplex bidirectional relay channel and provide inner and outer bounds for different relaying protocols in \cite{ray7}. Schnurr et al. derive an achievable rate region for the restricted two-way relay channel with partial decode-forward relaying in \cite{ray8}. El Gamal and Zahedi establish the capacity of the relay channel with orthogonal transmitting components which models frequency division in \cite{rorth}. Partial decode-forward relaying achieves its capacity, for which the half-duplex factor simplifies analysis. However, it is not always the case that half-duplex capacity is simpler or can be derived  directly from full-duplex capacity.

In addition, the block Markov coding structure in the full-duplex MAC-GF introduces dependency between contiguous codeblocks. As a result, backward decoding becomes the preferred technique to increase the rate region. However, backward decoding leads to excessive decoding delay. Fortunately in half-duplex systems, since each user cannot transmit and receive simultaneously, block Markov coding need not apply. Therefore, it may be the case that optimal half-duplex coding can be done independently for each codeblock, which removes the need for backward
decoding and the excessive decoding delay accompanied with it.

In this paper, we propose new coding schemes for the half-duplex MAC-GF, taking into account the half-duplex and short decoding delay constraints. Two main features allow the proposed schemes to meet these two constraints. First, these schemes perform the transmission in independent blocks without block Markovity. As a result, the destination can decode at the end of each block without any delay. Second, we use time division in each code block and divide each into three time slots. Allowing each user to either transmit or receive during each slot satisfies the half-duplex constraint. In the proposed schemes, each user alternatively transmits and receives during the first two time slots, then both transmit during the third one and the destination only decodes at the end of this time slot.

We consider two different coding schemes. Both schemes employ rate splitting and superposition encoding in each time slot. However, they have two main differences. First, in the first coding scheme, the codewords transmitted during the $3^{\text{rd}}$ time slot are superimposed  on the codewords of the first two. This is similar to the coding scheme of the full-duplex MAC-GF in \cite{fcc2005fof,cognitiveRT}. However, in the second coding scheme, these codewords are independent. Second, the first scheme employs partial decode-forward, while the second uses full decode-forward. For the Gaussian channel, we show that these two schemes achieve the same rate region and hence are equivalent.

For the first coding scheme, we also consider two different decoding techniques at the destination: separate decoding and joint decoding. In separate decoding, the destination starts from the $3^{\text{rd}}$ time slot and decodes the messages received in each time slot independently. In joint decoding, the destination uses signals received in all $3$ time slots to decode all messages simultaneously. Analysis as well as numerical results show that joint decoding strictly outperforms separate decoding by achieving a larger rate region.

We also derive two outer bounds for the half-duplex MAC-GF in a form similar to each achievable rate region
but on a larger input distribution. These outer bounds are derived using standard method that employs Fano's and data processing inequalities among multiple time slots. We also show that the second outer bound can be derived from the dependence balance outer bound for full-duplex MAC-GF in \cite{dbout}. Similar to the achievable rate regions, the two outer bounds are equivalent for the Gaussian channel. These bounds become tighter as the inter-user link qualities increase.
We also show that our outer bound becomes the capacity for the Gaussian physically degraded channel.
Finally, we extend our coding scheme, achievability, and outer bound to the $m$-user case.

This paper is organized as follows. Section \ref{sec:system_model} describes the half-duplex MAC-GF model. Section \ref{sec:achrr} presents the partial decode-forward based coding scheme and provides its achievable rate region for both joint and separate decoding techniques at the destination. Section \ref{sec: simpsch} presents a simplified decode-forward based scheme and its rate region. The outer bounds are provided in Section \ref{sec:out. cap}. Section \ref{sec:kachout} extends the channel to the $m$-user case and provides an achievability and outer bound. Application of the proposed coding schemes in the Gaussian channel is given in Section \ref{sec:cap. gau} with numerical results and comparison the outer bound. Finally, Section \ref{sec:conclusion} concludes the paper.


\section{Channel Model}\label{sec:system_model}
\begin{figure}[]
    \begin{center}
    \includegraphics[width=0.6\textwidth]{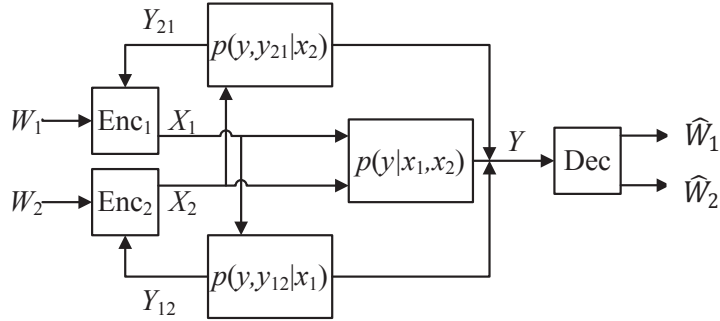}
    \caption{The half-duplex MAC-GF model.} \label{fig:system_model}
    \end{center}
\vspace*{-7mm}
\end{figure}

The two-user discrete memoryless half-duplex MAC-GF consists of two input alphabets ${\cal X}_1$
and ${\cal X}_2$, three output alphabets ${\cal Y}$, ${\cal Y}_{12}$, and ${\cal Y}_{21}$, and three conditional transition probabilities $p(y|x_1,x_2)$, $p(y,y_{12}|x_1)$, and $p(y,y_{21}|x_2)$ as shown in Fig.\ref{fig:system_model}. This channel is quite similar to the full-duplex MAC with generalized feedback as defined in \cite{fcc2005fof}. However, each user (the owner of the message $W_1$ or $W_2$) can only either be in transmit or receive mode but not in both. Hence, an additional requirement for the half-duplex MAC-GF is that no two transition probabilities occur at the same time. Because of this requirement, the coding scheme in \cite{fcc2005fof} can not be applied directly.

A $(\lceil2^{nR_1}\rceil,\lceil2^{nR_2}\rceil,n)$ code for this channel consists of two message sets $W_1=\{1,\ldots,\lceil2^{nR_1}\rceil\}$ and $W_2=\{1,\ldots,\lceil2^{nR_2}\rceil\}$, two encoding functions $f_{1i},f_{2i},\;i=1,\ldots,n$, and one decoding function $g$ defined as
\noindent
\begin{align}
&f_{1i}: W_1\times {\cal Y}_{21}^{i-1}\rightarrow{\cal X}_1,\; {}i=1,\ldots,n \nonumber\\
&f_{2i}: W_2\times {\cal Y}_{12}^{i-1}\rightarrow{\cal X}_2,\; {}i=1,\ldots,n \nonumber\\
&g: {\cal Y}^n \rightarrow W_1\times W_2.
\label{Eq:fR}
\end{align}
\noindent Finally, $P_e$ is the average error probability defined as $P_e=P(g(Y^n)\neq (W_1,W_2))$. A rate pair $(R_1,R_2)$ is said to be achievable if there
exists a $(\lceil2^{nR_1}\rceil,\lceil2^{nR_2}\rceil,n)$ code such that $P_e\rightarrow0$ as $n\rightarrow\infty$. The capacity region is the closure of the set of all achievable rates
$(R_1,R_2)$.

The three transition probabilities of the half-duplex MAC-GF can be modeled using time division such that for $n$ uses of the channel, the transition probability $p(y,y_{12},y_{21}|x_1,x_2)$ can be expressed as
\noindent
\begin{align}\label{tdchm}
P^{\bullet}=p(y,y_{12},y_{21}|x_1,x_1)=&\;p(y,y_{12}|x_1)\left(u(k)-u(k-\alpha_1n)\right)\nonumber\\
&+\;p(y,y_{21}|x_2)\left(u(k-\alpha_1n)-u(k-(\alpha_1+\alpha_2)n)\right)\nonumber\\
&+\;p(y|x_1,x_2)\left(u(k-(\alpha_1+\alpha_2)n)-u(k-n)\right)
\end{align}
\noindent where $0\leq \alpha_1+\alpha_2 \leq 1$ and $u(n)$ is the discrete-time unit step function.

Thus, each transmission block is divided into three time slots with variable durations $\alpha_1,$ $\alpha_2$ and
$(1-\alpha_1-\alpha_2)$. While the destination is always in receiving mode, each user either transmits or receives during
the first two time slots and both of them transmit during the third slot. Specifically, as in
Fig.\ref{fig:system_model}, the channels $p(y,y_{12}|x_1),$ $p(y,y_{21}|x_2)$ and $p(y|x_1,x_2)$ occur during the
$1^{\text{st}},$ $2^{\text{nd}}$ and $3^{\text{rd}}$ time slots, respectively. For clarity, instead of always using
$y$ for the channel out at the destination, we also refer to it in each time slot differently as $y_1, y_2$ or $y_3$.
\section{Coding Schemes with Partial Decode-Forward (PDF) Relaying}\label{sec:achrr}
In this section, we propose and analyze a coding scheme for the half-duplex MAC-GF based on partial decode-forward (PDF). The transmission is done in independent blocks of length $n$.
Each user employs rate splitting and superposition coding in each time slot. Consider the first user; it splits its message, $W_1$, into three parts $(W_{10},W_{12},W_{13})$ where $(W_{10},W_{13})$ are the private parts and $W_{12}$ is the public part. While the private parts are transmitted directly to the destination at rates $R_{10}$ and $R_{13}$, respectively, the public part is transmitted to
the destination in cooperation with the second user at rate $R_{12}$. During the $1^{\text{st}}$ time slot, the first user sends $(W_{10},W_{12})$, while during the $3^{\text{rd}}$ time slot, it sends $(W_{21},W_{12},W_{13})$. The transmission of the second user is similar. In this coding scheme, the codewords of the private parts $(W_{23},W_{13})$ sent during the $3^{\text{rd}}$ time slot are superimposed on the codewords of the cooperative parts during the first two time slots, which makes these codewords dependent. We name this scheme the PDF scheme and show that the following rate region is achievable.\\\\

\textbf{Theorem $1$.\;}\emph{The achievable rate region for the half-duplex MAC-GF using the PDF scheme with full decoding at each user and \textbf{joint} decoding at the destination is the convex closure of all rate pairs $(R_1,R_2)$ satisfying}
\begin{align}\label{th1rr}
R_1\leq&\; \alpha_1I(X_{10};Y_{12})+\alpha_3 I(X_{13};Y_3|X_{23},U,V)\nonumber\\
R_2\leq&\; \alpha_2I(X_{20};Y_{21})+\alpha_3 I(X_{23};Y_3|X_{13},U,V)\nonumber\\
R_1+R_2\leq&\; \alpha_1I(X_{10};Y_{12})+\alpha_2I(X_{20};Y_{21})+\alpha_3 I(X_{13},X_{23};Y_3|U,V)\nonumber\\
R_1+R_2\leq&\; \alpha_1I(X_{10};Y_1)+\alpha_2I(X_{20};Y_{21})+\alpha_3 I(X_{13},X_{23};Y_3|V)\nonumber\\
R_1+R_2\leq&\; \alpha_1I(X_{10};Y_{12})+\alpha_2I(X_{20};Y_2)+\alpha_3 I(X_{13},X_{23};Y_3|U)\nonumber\\
R_1+R_2\leq&\; \alpha_1I(X_{10};Y_1)+\alpha_2I(X_{20};Y_2)+\alpha_3 I(X_{13},X_{23};Y_3)
\end{align}
\noindent \emph{for some joint distribution that factors as}
\begin{align}\label{ipdis}
P^{\ast}=p(x_{10},u)p(x_{20},v)p(x_{13}|u,v)p(x_{23}|u,v)
\end{align}
\noindent \emph{where $\alpha_3=(1-\alpha_1-\alpha_2)$.}

Next, we provide a full description of the encoding and decoding technique that achieves the above region,
with the summary in Table I.
\subsection{Coding Scheme}\label{costh1}
As mentioned previously, the encoding is performed using rate splitting and superposition encoding.
The codebook generation is done as follows.
\subsubsection{Codebook generation}
Fix $P^{\ast}$ in (\ref{ipdis}) and generate
\noindent
\begin{itemize}
\item
$2^{nR_{12}}$ i.i.d sequences $u^{n}(w_{12})\sim\prod_{i=1}^np(u_{i})$,
\item
$2^{nR_{21}}$ i.i.d sequences $v^{n}(w_{21})\sim\prod_{i=1}^np(v_{i})$.
\end{itemize}
\noindent Then for each $u^{n}(w_{12})$ and each $v^{n}(w_{21})$, generate
\noindent
\begin{itemize}
\item
$2^{nR_{10}}$ i.i.d sequences $x_{10}^{n}(w_{10},w_{12})\sim\prod_{i=1}^np(x_{10i}|u_i),$
\item
$2^{nR_{20}}$ i.i.d sequences $x_{20}^{n}(w_{20},w_{21})\sim\prod_{i=1}^np(x_{20i}|v_i)$.
\end{itemize}
Finally, for each pair $(u^{n}(w_{12}),v^{n}(w_{21}))$, generate
\noindent
\begin{itemize}
\item
$2^{nR_{13}}$ i.i.d sequences $x_{13}^{n}(w_{13},w_{12},w_{21})\sim\prod_{i=1}^np(x_{13i}|u_i,v_i),$
\item
$2^{nR_{23}}$ i.i.d sequences $x_{23}^{n}(w_{23},w_{12},w_{21})\sim\prod_{i=1}^np(x_{23i}|u_i,v_i).$
\end{itemize}
\begin{figure*}[t]
\normalsize
\begin{center}
\begin{tabular}{|c|c|c|c|}
\hline
{}&$1^\text{st}$ slot with length $\alpha_1 n$ &$2^\text{nd}$ slot with length $\alpha_2 n$&$3^\text{rd}$ slot with length $(1-\alpha_1-\alpha_2)n$\\
\hline
first user&$x_{10}^{\alpha_1 n}(w_{10},w_{12})$&$--$&$x_{13, (\alpha_1+\alpha_2) n +1}^{n}(w_{13},w_{12},\tilde{w}_{21})$\\
\hline
second user&$--$&$x_{20}^{\alpha_2 n}(w_{20},w_{12})$&$x_{23, (\alpha_1+\alpha_2) n +1}^{n}(w_{23},\tilde{w}_{12},w_{21})$\\
\hline
\multirow {2}{*}{$Y_{21}$}&$--$&$(\tilde{w}_{20},\tilde{w}_{21})$ (full Dec.)&$--$\\
\cline{2-4}
&$--$&$\tilde{w}_{21}$ (partial Dec.)&$--$\\
\hline
\multirow {2}{*}{$Y_{12}$}&$(\tilde{w}_{10},\tilde{w}_{12})$ (full Dec.)&$--$&$--$\\
\cline{2-4}
&$\tilde{w}_{12}$ (partial Dec.)&$--$&$--$\\
\hline
$Y$&$Y_1$&$Y_2$&$Y_3$\\
\hline
Sep. Dec. &$\hat{w}_{10}$&$\hat{w}_{20}$&$\leftarrow(\hat{w}_{12},\hat{w}_{21},\hat{w}_{13},\hat{w}_{23})$\\
\hline
Joint Dec.& \multicolumn{3}{c|}{$(\hat{w}_{12},\hat{w}_{21},\hat{w}_{10},\hat{w}_{20},\hat{w}_{13},\hat{w}_{23})$}\\
\hline
\end{tabular}
\\
\vspace*{2mm}
{\small Table I: The encoding and decoding techniques for the PDF scheme.}\\
\vspace*{2mm}
\end{center}
\vspace*{-11mm}
\end{figure*}

\subsubsection{Encoding}
In order to send the message pair $(w_1,w_2)$, the first user sends $x_{10}^{\alpha_1 n}(w_{10},w_{12})$ during the $1^\text{st}$ time slot, while
the second user sends $x_{20}^{\alpha_2 n}(w_{20},w_{12})$ during the $2^\text{nd}$ time slot. At the end of the $1^\text{st}$ and $2^\text{nd}$ time slots, the second user and the first user will have the estimated values $(\tilde{w}_{10},\tilde{w}_{12})$ and $(\tilde{w}_{20},\tilde{w}_{21})$, respectively. Then, the first user
sends $x_{13, (\alpha_1+\alpha_2) n +1}^{n}(w_{13},w_{12},\tilde{w}_{21})$ and the second user sends $x_{23, (\alpha_1+\alpha_2)n +1}^{n}(w_{23},\tilde{w}_{12},w_{21})$ during the last time slot. Hence cooperation occurs via decode-forward relaying. Each user decodes the other user's messages during the first two time slots, then forwards the public part of this message during the third time slot, in addition to another private message to the destination.

In this scheme, the codewords of the private parts $(w_{13},w_{23})$ are superimposed on the codewords of the public parts, which is similar to the full-duplex coding scheme in \cite{fcc2005fof,cognitiveRT}. However, the codewords in the proposed scheme encode messages in the same block, while in \cite{fcc2005fof,cognitiveRT}, the codewords for private parts are superimposed on those of the  public parts in the previous block. As a result, the proposed scheme has independent transmission blocks while the coding scheme in \cite{fcc2005fof,cognitiveRT} has Markov dependent blocks. Because of independent blocks, decoding at the destination can be done at the end of each block. Hence, there is no decoding delay in the proposed scheme, while in \cite{fcc2005fof,cognitiveRT}, there is a long delay resulting from backward decoding. Also, we can see that the proposed scheme includes the classical MAC and the classical TDMA scheme as special cases  when $\alpha_1=\alpha_2=0$ and  $\alpha_1=\alpha_2=0.5$, respectively.

\subsubsection{Decoding Technique}
\subsubsection*{Full Decoding at each user}
At the end of the $1^{\text{st}}$ time slot, the second user applies joint typicality rule to decode both message parts  $(w_{10},w_{12})$ from its received sequence $\boldsymbol{Y\!}_{12}$. Specifically, the second user looks for a unique message pair $(\tilde{w}_{10},\tilde{w}_{12})$ that satisfies
\noindent
\begin{align*}
(u^{\alpha_1n}(\tilde{w}_{12}),x_{10}^{\alpha_1 n}(\tilde{w}_{10},\tilde{w}_{12}),\boldsymbol{Y\!}_{12})\in A_\epsilon ^{\alpha_1n}.
\end{align*}
\noindent Similarly, the first user decodes the unique message pair $(\tilde{w}_{20},\tilde{w}_{21})$ such that
\noindent
\begin{align*}
(v^{\alpha_2n}(\tilde{w}_{21}),x_{20}^{\alpha_2 n}(\tilde{w}_{20},\tilde{w}_{21}),\boldsymbol{Y\!}_{21})\in A_\epsilon ^{\alpha_2n}.
\end{align*}
Following standard joint typicality analysis as in \cite{hansen1977mfr}, the error probabilities in these decoding go to zero as $n\rightarrow\infty$ if the following rate constraints are satisfied:
\noindent
\begin{align}\label{Eq:datdes}
R_{10}\leq&\;\alpha_1 I(X_{10};Y_{12}|U)=I_1\nonumber\\
R_{10}+R_{12}\leq&\; \alpha_1 I(X_{10};Y_{12})=I_2\nonumber\\
R_{20}\leq&\;\alpha_2 I(X_{20};Y_{21}|V)=I_3\nonumber\\
R_{20}+R_{21}\leq&\; \alpha_2 I(X_{20};Y_{21})=I_4
\end{align}
\subsubsection*{Joint Decoding at the destination}
From Table I, we can see that all received signals at the destination $(\boldsymbol{Y_1, Y_2, Y_3})$ include information about the cooperative message parts $(w_{12}, w_{21})$. Thus, joint decoding among these three signals will improve the rate region.

In joint decoding, the destination decodes the message vector $(\hat{w}_{12},\hat{w}_{21},\hat{w}_{10},\hat{w}_{20},\hat{w}_{13},\hat{w}_{23})$ using joint typicality \cite{hansen1977mfr} or joint ML \cite{abramowitz1972hmf} decoding based on the whole received sequence $\boldsymbol{y}=(y_1^{\alpha_1 n} y_2^{\alpha_2 n}$ $y_3^{\alpha_3n})$. Specifically, the destination looks for a unique message vector $(\hat{w}_{12},\hat{w}_{21},\hat{w}_{10},\hat{w}_{20},\hat{w}_{13},\hat{w}_{23})$ that simultaneously satisfies

\noindent
\begin{align}\label{Eq:ddest}
\{(u^{\alpha_1n}(\hat{w}_{12}),x_{10}^{\alpha_1 n}(\hat{w}_{10},\hat{w}_{12}),\boldsymbol{Y\!}_1)\in& A_\epsilon ^{\alpha_1n},\nonumber\\
\text{and}\;(v^{\alpha_2n}(\hat{w}_{21}),x_{20}^{\alpha_2 n}(\hat{w}_{20},\hat{w}_{21}),\boldsymbol{Y\!}_{2})\in& A_\epsilon ^{\alpha_2n}, \nonumber\\
\text{and}\;(u^{\alpha_3n}(\hat{w}_{12}),v^{\alpha_3n}(\hat{w}_{21}),x_{13}^{\alpha_3 n}(\hat{w}_{13},\hat{w}_{12},\hat{w}_{21}),x_{23}^{\alpha_3 n}(\hat{w}_{23},\hat{w}_{12},\hat{w}_{21}),\boldsymbol{Y\!}_{3})\in& A_\epsilon ^{\alpha_3n}\}.
\end{align}
The error analysis of this decoding technique is given in Appendix A and it leads to the following rate constraints:
\noindent
\begin{align}\label{Eq:d2dec}
R_{10}\leq&\;\alpha_1 I(X_{10};Y_{1}|U)=J_1 \nonumber\\
R_{20}\leq&\;\alpha_2 I(X_{20};Y_{2}|V)=J_2\nonumber \\
R_{13}\leq&\;\alpha_3 I(X_{13};Y_3|U,V,X_{23})=J_3\nonumber \\
R_{23}\leq&\;\alpha_3 I(X_{23};Y_3|U,V,X_{13})=J_4\nonumber \\
R_{13}+R_{23}\leq&\;\alpha_3 I(X_{13},X_{23};Y_3|U,V)=J_5\nonumber\\
R_{1}+R_{23}\leq&\;\alpha_1 I(X_{10};Y_1)+\alpha_3 I(X_{13},X_{23};Y_3|V)=J_6\nonumber\\
R_{2}+R_{13}\leq&\;\alpha_2 I(X_{20};Y_2)+\alpha_3 I(X_{13},X_{23};Y_3|U)=J_7\nonumber\\
R_1+R_2\leq&\; \alpha_1 I(X_{10};Y_1)+\alpha_2 I(X_{20};Y_2)+\alpha_3 I(X_{13},X_{23};Y_3)=J_8.
\end{align}
The same rate constraints can be obtained using joint ML decoding as shown in \cite{haivu3}.
Now, by applying Fourier-Motzkin Elimination (FME) to the inequalities in (\ref{Eq:d2dec}) and (\ref{Eq:datdes}), the achievable rates in terms of $R_1=R_{10}+R_{12}+R_{13}$ and $R_2=R_{20}+R_{21}+R_{23}$ can be expressed as in Theorem 1.

\subsubsection*{Remark 1: Alternative partial decoding at each user}\label{secpdu}
In the above scheme, each user decodes both the public and private message parts of the other user. However, the private part is not forwarded. Thus, alternatively,
each user can also perform partial decoding. We will show, however, that partial decoding is of no advantage for Gaussian channels.

In partial decoding, each user decodes only the public part of the other user during the first two time slots. Specifically, the second user decodes $w_{12}$ from $\boldsymbol{Y\!}_{12}$ by looking for a unique $\tilde{w}_{12}$ such that $(u^{\alpha_1n}(\tilde{w}_{12}),\boldsymbol{Y\!}_{12})\in A_\epsilon ^{\alpha_1n}.$ Similarly, the first user decodes $w_{21}$ from $\boldsymbol{Y\!}_{21}$ by looking for a unique $\tilde{w}_{21}$ such that $(v^{\alpha_1n}(\tilde{w}_{21}),\boldsymbol{Y\!}_{21})\in A_\epsilon ^{\alpha_2n}.$ Again, by applying joint typicality analysis, the probability of error goes to zero as $n\rightarrow \infty$ if the following two constraints satisfied:
\begin{align*}
R_{12}\leq&\; \alpha_1 I(U;Y_{12})\\
R_{21}\leq&\; \alpha_2 I(V;Y_{21}).
\end{align*}
\noindent By applying FME for these new constraints together with the constraints in (\ref{Eq:d2dec}), we get a rate region region similar to Theorem 1, except that we need to replace
\begin{align*}
&I(X_{10};Y_{12})\; \text{by}\; I(U;Y_{12})+I(X_{10};Y_1|U), \\
\text{and}\;&I(X_{20};Y_{21})\; \text{by}\; I(V;Y_{21})+I(X_{20};Y_2|V).
\end{align*}
\noindent To compare between the two regions, note that in Theorem 1, we have
\begin{align*}
&I(X_{10};Y_{12})=I(U,X_{10};Y_{12})=I(U;Y_{12})+I(X_{10};Y_{12}|U),\\
\text{and}\;&I(X_{20};Y_{21})=I(V,X_{20};Y_{21})=I(V;Y_{21})+I(X_{20};Y_{21}|V).
\end{align*}
Therefore, we just need to compare the pairs $(I(X_{10};Y_{12}|U), I(X_{10};Y_1|U))$
and $(I(X_{20};Y_{21}|V), I(X_{20};Y_2|V))$. If $I(X_{10};Y_{12}|U)>I(X_{10};Y_1|U)$ and
$I(X_{20};Y_{21}|V)>I(X_{20};Y_2|V)$, then full decoding at each user leads to a larger rate region for a fixed input distribution.
This is the case for Gaussian channel explained in Section \ref{sec:cap. gau} when the inter-user links are
stronger than direct links.

\subsection{Alternative separate decoding at the destination}
In order to recognize the efficacy of joint decoding in enlarging the rate region, we compare this decoding technique with separate decoding, in which the destination performs decoding separately in each time slot as shown in Table I. It decodes starting at the $3^{\text{rd}}$ time slot, then goes back to the $2^{\text{nd}}$ and $1^{\text{st}}$ ones.

In separate decoding, starting from the $3^{\text{rd}}$ time slot, the destination uses the received signal $\boldsymbol{Y\!}_{3}$ to decode the message vector $(\hat{w}_{12},\hat{w}_{21},\hat{w}_{13},\hat{w}_{23})$ using joint typicality decoding. In this time slot, the channel looks like a MAC with common message \cite{haykin2005crb} and the destination looks for a unique message vector $(\hat{w}_{12},\hat{w}_{21},\hat{w}_{13},\hat{w}_{23})$ that satisfies
\begin{align*}
(u^{\alpha_3n}(\hat{w}_{12}),v^{\alpha_3n}(\hat{w}_{21}),x_{13}^{\alpha_3 n}(\hat{w}_{13},\hat{w}_{12},\hat{w}_{21}),x_{23}^{\alpha_3 n}(\hat{w}_{23},\hat{w}_{12},\hat{w}_{21}),\boldsymbol{Y\!}_{3})\in A_\epsilon ^{\alpha_3n}.
\end{align*}
\noindent Then, the destination goes back to the first $2$ time slots, in which the channel looks like a broadcast channel with superposition coding \cite{hansen1977mfr}. It uses the received signals $\boldsymbol{Y\!}_{1}$ $(\boldsymbol{Y\!}_{2})$ to decode $\hat{w}_{10}$ $(\hat{w}_{20})$, respectively, assuming that it already decodes the $(\hat{w}_{12},\hat{w}_{21})$ correctly in the $3^{\text{rd}}$ time slot. Specifically, the destination looks for a unique message $\hat{w}_{10}$ $(\hat{w}_{20})$ which satisfies
\begin{align*}
(u^{\alpha_1n}(\hat{W}_{12}),x_{10}^{\alpha_1 n}(\hat{w}_{10},\hat{W}_{12}),\boldsymbol{Y\!}_1)\in A_\epsilon ^{\alpha_1n}
\end{align*}
\noindent or
\begin{align*}
(v^{\alpha_2n}(\hat{W}_{21}),x_{20}^{\alpha_2 n}(\hat{w}_{20},\hat{W}_{21}),\boldsymbol{Y\!}_{2})\in A_\epsilon ^{\alpha_2n}.
\end{align*}
\noindent for a given $(\hat{w}_{12},\hat{w}_{21})$. Following the analysis in \cite{hansen1977mfr,haykin2005crb}, the rate constraints that make the probabilities of error to go to zero as $n\rightarrow \infty$ are as follows.
\noindent
\begin{align}\label{Eq:d1dec}
R_{10}\leq&\;\alpha_1 I(X_{10};Y_{1}|U)=I_5\nonumber\\
R_{20}\leq&\;\alpha_2 I(X_{20};Y_{2}|V)=I_6\nonumber\\
R_{13}\leq&\;\alpha_3 I(X_{13};Y_3|U,V,X_{23})=I_7\nonumber\\
R_{23}\leq&\;\alpha_3 I(X_{23};Y_3|U,V,X_{13})=I_8\nonumber\\
R_{13}+R_{23}\leq&\;\alpha_3 I(X_{13},X_{23};Y_3|U,V)=I_9\nonumber\\
R_{12}+R_{13}+R_{23}\leq&\;\alpha_3 I(X_{13},X_{23};Y_3|V)=I_{10}\nonumber\\
R_{21}+R_{13}+R_{23}\leq&\;\alpha_3 I(X_{13},X_{23};Y_3|U)=I_{11}\nonumber\\
R_{12}+R_{21}+R_{13}+R_{23}\leq&\alpha_3 I(X_{13},X_{23};Y_3)=I_{12}.
\end{align}
\noindent The derivation of these rate constraints are given in Appendix B.

By combining these constraints together with the constraints of full decoding at each user in (\ref{Eq:datdes}), we get the following corollary.

\textbf{Corollary $1$.} \emph{The achievable rate region for the half-duplex MAC-GF using the PDF scheme with full decoding at each user and \textbf{separate} decoding at the destination is the convex closure of all rate pairs $(R_1,R_2)$ satisfying}
\noindent
\begin{align}\label{co2rr}
R_1\leq&\; \alpha_1I(X_{10};Y_{12})+\alpha_3 I(X_{13};Y_3|X_{23},U,V)\\
R_2\leq&\; \alpha_2I(X_{20};Y_{21})+\alpha_3 I(X_{23};Y_3|X_{13},U,V)\nonumber\\
R_1+R_2\leq&\; \alpha_1I(X_{10};Y_{12})+\alpha_2I(X_{20};Y_{21})+\alpha_3 I(X_{13},X_{23};Y_3|U,V)\nonumber\\
R_1+R_2\leq&\; \alpha_1 \text{min}(I(X_{10};Y_{12}|U),\;I(X_{10};Y_{1}|U))+\alpha_2I(X_{20};Y_{21})+\alpha_3 I(X_{13},X_{23};Y_3|V)\nonumber\\
R_1+R_2\leq&\; \alpha_2\text{min}(I(X_{20};Y_{21}|V),\;I(X_{20};Y_{2}|V))+\alpha_1I(X_{10};Y_{12})+\alpha_3 I(X_{13},X_{23};Y_3|U)\nonumber\\
R_1+R_2\leq&\;\alpha_1 \text{min}(I(X_{10};Y_{12}|U),\;I(X_{10};Y_{1}|U))+\alpha_2\text{min}(I(X_{20};Y_{21}|V),\;I(X_{20};Y_{2}|V))+\alpha_3 I(X_{13},X_{23};Y_3)\nonumber
\end{align}
\emph{for some joint distribution $P^{\ast}=p(x_{10},u)p(x_{20},v)p(x_{13}|u,v)p(x_{23}|u,v)$ and $0\leq \alpha_1+\alpha_2 \leq 1$.}

\subsubsection*{Remark $2$} Although the first $3$ constraints in this rate region are the same as those in Theorem 1, the other $3$ are smaller. This is because $(I(X_{10};Y_{1})\geq I(X_{10};Y_{1}|U))$ and $(I(X_{20};Y_{2})\geq I(X_{20};Y_{2}|V))$. Hence, this rate region will always be smaller than that in Theorem 1.

A simple explanation for this difference between the two regions is as follows.  In separate decoding, the destination only uses the received signal in the $3^{\text{rd}}$ time slot to decode the public parts $(w_{12},w_{21})$ but ignores the received signals in the first two time slots, even though they include information about the public parts. On the other hand, in joint decoding, the destination uses the received signals from the all $3$ time slots to  decode the transmitted messages $(W_1,W_2)$. As a result, the set of error events in separate decoding will be bigger than in joint decoding and hence, the rate region for separate decoding is smaller.

Consequently, we can see that although the channel capacity maybe known in each time slot as of the (degraded) broadcast channel \cite{hansen1977mfr} or the MAC with common message \cite{haykin2005crb}, the capacity for the half-duplex channel cannot be simply derived from these existing capacities. The half-duplex capacity is still an open problem and in Section \ref{sec:out. cap}, we provide an outer bound for it.

\subsubsection*{Remark $3$} The rate region in Theorem 1 was obtained by applying FME for (\ref{Eq:datdes}) and (\ref{Eq:d2dec}). However, when applying FME, the rate constraints involving $(I_1, I_3)$ and $(J_1, J_2)$ given in (\ref{Eq:datdes}) and (\ref{Eq:d2dec}) appeared to be redundant and did not affect the rate region in Theorem 1. Any
value for $(R_{10}, R_{20})$ will not affect the rate region. Therefore, there is no need for the private part $(w_{10},w_{20})$ in the first two time slots if joint decoding is used at the destination (note however that $(w_{10},w_{20})$ is still necessary with separate decoding). Next, we propose a new scheme taking this remark into account.

\section{Simplified Decode-Forward (DF) Scheme}\label{sec: simpsch}
In this section, we provide a simplified coding scheme in which each user splits it message into only two parts, $w_1=(w_{12},w_{13})$ and $w_2=(w_{21},w_{23})$. The transmission in each time slot is similar to the previous scheme but without $(w_{10},w_{20})$. Besides, the codeword in the $3^{\text{rd}}$ time slot is independent from codewords in the first two, instead of being superimposed on them as before.\\\\\\

\textbf{Theorem $2$.} \emph{The achievable rate region for the half-duplex MAC-GF using the DF with joint decoding at the destination can be expressed as the convex closure of all rate pairs $(R_1,R_2)$ satisfying}
\begin{align}\label{sth1rr1}
R_1\leq&\; \alpha_1I(X_{12};Y_{12})+\alpha_3 I(X_{13};Y_3|X_{23},S)\nonumber\\
R_2\leq&\; \alpha_2I(X_{21};Y_{21})+\alpha_3 I(X_{23};Y_3|X_{13},S)\nonumber\\
R_1+R_2\leq&\; \alpha_1I(X_{12};Y_{12})+\alpha_2I(X_{21};Y_{21})+\alpha_3 I(X_{13},X_{23};Y_3|S)\nonumber\\
R_1+R_2\leq&\; \alpha_1I(X_{12};Y_1)+\alpha_2I(X_{21};Y_{21})+\alpha_3 I(X_{13},X_{23};Y_3)\nonumber\\
R_1+R_2\leq&\; \alpha_1I(X_{12};Y_{12})+\alpha_2I(X_{21};Y_2)+\alpha_3 I(X_{13},X_{23};Y_3)\nonumber\\
R_1+R_2\leq&\; \alpha_1I(X_{12};Y_1)+\alpha_2I(X_{21};Y_2)+\alpha_3 I(X_{13},X_{23};Y_3)
\end{align}
\noindent \emph{for some $p(x_{12})p(x_{21})p(s)p(x_{13}|s)p(x_{23}|s)$, where $\alpha_1+\alpha_2+\alpha_3=1$}
\begin{proof}
We provide the encoding and decoding techniques for this new scheme in which each user splits its message into only two parts $w_1=(w_{12},w_{13})$ and $w_2=(w_{21},w_{23})$.

\subsubsection{Codebook generation}
Fix $P_s^{\ast}=p(x_{12})p(x_{21})p(u)p(x_{13}|u)p(x_{23}|u)$, then generate
\noindent
\begin{itemize}
\item
$2^{nR_{12}}$ i.i.d sequences $x_{12}^{n}(w_{12})\sim\prod_{i=1}^np(x_{12i})$
\item
$2^{nR_{21}}$ i.i.d sequences $x_{21}^{n}(w_{21})\sim\prod_{i=1}^np(x_{21i})$
\item
$2^{n(R_{12}+R_{21})}$ i.i.d sequences $s^{n}(w_{12},w_{21})\sim\prod_{i=1}^np(s_{i})$.
\end{itemize}
\noindent Then for each pair $s^n(w_{12},w_{21})$, generate
\noindent
\begin{itemize}
\item
$2^{nR_{13}}$ i.i.d sequences $x_{13}^{n}(w_{13},w_{12},w_{21})\sim\prod_{i=1}^np(x_{13i}|s_i)$
\item
$2^{nR_{23}}$ i.i.d sequences $x_{23}^{n}(w_{23},w_{12},w_{21})\sim\prod_{i=1}^np(x_{23i}|s_i)$.
\end{itemize}
\noindent The encoding and decoding of this scheme is similar to the PDF scheme of Theorem 1 but without $(w_{10},w_{20})$. Moreover, codewords $\left(x_{13}^{n}(w_{13},w_{12},w_{21}),x_{23}^{n}(w_{23},w_{12},w_{21})\right)$ are superimposed on codeword $s^n(w_{12},w_{21})$, which even though encodes the same message pairs $(w_{12},w_{21})$ as codewords $(x_{12}^{n}(w_{12}),x_{21}^{n}(w_{21}))$, is independent from them because it is generated according to an independent distribution.
\subsubsection{Encoding}
For the two users to send the message pair $(w_1,w_2)$, the first and the second users transmit $x_{12}^{\alpha_1n}(w_{12})$ and $x_{21}^{\alpha_2n}(w_{21})$ during the $1^{\text{st}}$ and the $2^{\text{nd}}$ time slots, respectively. They also decode $\tilde{w}_{21}$ and $\tilde{w}_{12}$ at the end of these two time slots. Then, during the last time slots, they send $x_{13, (\alpha_1+\alpha_2) n +1}^{n}(w_{13},w_{12},\tilde{w}_{21})$ and $x_{23, (\alpha_1+\alpha_2)n +1}^{n}(w_{23},\tilde{w}_{12},w_{21})$.
\subsubsection{Decoding}
\subsubsection*{At each user}
At the end of the $1^{\text{st}}$ time slot, the second user decodes $w_{12}$ by looking for a unique $\tilde{w}_{12}$ that satisfies
\noindent
\begin{align*}
(x_{12}^{\alpha_1n}(\tilde{w}_{12}),\boldsymbol{Y\!}_{12})\in A_\epsilon ^{\alpha_1n}.
\end{align*}
\noindent Similarly, the first user decodes $w_{21}$ by looking for a unique $\tilde{w}_{21}$ that satisfies
\noindent
\begin{align*}
(x_{21}^{\alpha_2n}(\tilde{w}_{21}),\boldsymbol{Y\!}_{21})\in A_\epsilon ^{\alpha_2n}.
\end{align*}
Following standard joint typicality analysis \cite{hansen1977mfr}, we obtain the following rate constraints:
\noindent
\begin{align}\label{Eq:sdecuser}
R_{12}\leq&\;\alpha_1 I(X_{12};Y_{12})=H_1\nonumber\\
R_{21}\leq&\;\alpha_2 I(X_{21};Y_{21})=H_2
\end{align}
\subsubsection*{Joint decoding at the destination}
Similar to the joint decoding in Section \ref{costh1}, the destination utilizes the received sequence from all $3$ time slots to decode the transmitted messages by looking for a unique message vector $(\hat{w}_{12},\hat{w}_{21},\hat{w}_{13},\hat{w}_{23})$ that simultaneously satisfies
\begin{align}\label{sdecdes}
\{(x_{12}^{\alpha_1n}(\hat{w}_{12}),\boldsymbol{Y\!}_1)\in& A_\epsilon ^{\alpha_1n},\nonumber\\
\text{and}\; (x_{21}^{\alpha_2n}(\hat{w}_{21}),\boldsymbol{Y\!}_{2})\in& A_\epsilon ^{\alpha_2n}, \nonumber\\
\text{and}\; (s^{\alpha_3n}(\hat{w}_{12},\hat{w}_{21}),x_{13}^{\alpha_3 n}(\hat{w}_{13},\hat{w}_{12},\hat{w}_{21}),x_{23}^{\alpha_3 n}(\hat{w}_{23},\hat{w}_{12},\hat{w}_{21}),\boldsymbol{Y\!}_{3})\in& A_\epsilon ^{\alpha_3n}\}.
\end{align}
The error analysis of this decoding technique is similar to that in Section \ref{costh1} and leads to the following rate constraints:
\noindent
\begin{align}\label{srdecdes}
R_{13}\leq&\;\alpha_3 I(X_{13};Y_3|S,X_{23})=H_3\nonumber \\
R_{23}\leq&\;\alpha_3 I(X_{23};Y_3|S,X_{13})=H_4\nonumber \\
R_{13}+R_{23}\leq&\;\alpha_3 I(X_{13},X_{23};Y_3|S)=H_5\nonumber\\
R_{1}+R_{23}\leq&\;\alpha_1 I(X_{12};Y_1)+\alpha_3 I(X_{13},X_{23};Y_3)=H_6\nonumber\\
R_{2}+R_{13}\leq&\;\alpha_2 I(X_{21};Y_2)+\alpha_3 I(X_{13},X_{23};Y_3)=H_7\nonumber\\
R_1+R_2\leq&\; \alpha_1 I(X_{12};Y_1)+\alpha_2 I(X_{21};Y_2)+\alpha_3 I(X_{13},X_{23};Y_3)=J_8
\end{align}
\noindent By applying Fourier-Motzkin Elimination (FME) to the inequalities in (\ref{Eq:sdecuser}) and (\ref{srdecdes}),  we get the achievable rates in terms of $R_1=R_{12}+R_{13}$ and $R_2=R_{21}+R_{23}$ as in Theorem 2.
\end{proof}
\subsubsection*{Remark $4$} In (\ref{sth1rr1}), the second and the third  constraints on the sum rate become redundant if $I(X_{12};Y_{12})>I(X_{12};Y_1)$ and $I(X_{21};Y_{21})>I(X_{21};Y_2)$.

\subsubsection*{Remark $5$} Although this rate region looks slightly different from that of the PDF scheme given in Theorem 1, Appendix E shows that the two regions are equivalent for Gaussian channels. For the discrete memoryless channel (DMC), this equivalency may not hold in general.
\section{Outer Bounds}\label{sec:out. cap}
In this section, we provide an outer bound with rate constraints similar to the PDF scheme and another one similar to the DF scheme. During the third time slot, the channel looks like the MAC with common message while during the first two time slots, it looks like a broadcast channel. Although the capacity is known for the MAC with common message \cite{haykin2005crb}, it is not known in general for the broadcast channel. Furthermore, the MAC-GF encompasses as a special case the relay channel of which the capacity is also not known in general. Next, we provide outer bounds to the half-duplex MAC-GF in a form similar to the achievable regions. These bounds are tight as the inter-user links are noticeably better than the link between each user and the destination.
\subsection{An Outer Bound Similar to the PDF Region}
Using standard Fano's and data processing inequalities among three time slots, we show in Appendix C that an outer bound for the half-duplex MAC-GF can be expressed as in the following theorem.\\\\\\

\textbf{Theorem $3$.} \emph{An outer bound of the half-duplex MAC-GF consists of the convex hull of all rate pairs $(R_1,R_2)$ satisfying }
\noindent
\begin{align}\label{outeq1}
R_1\leq&\; \alpha_1I(X_{10};Y_1,Y_{12})+\alpha_3 I(X_{13};Y_3|X_{23},U,V)\nonumber\\
R_2\leq&\; \alpha_2I(X_{20};Y_2,Y_{21})+\alpha_3 I(X_{23};Y_3|X_{13},U,V)\nonumber\\
R_1+R_2\leq&\; \alpha_1I(X_{10};Y_1,Y_{12})+\alpha_2I(X_{20};Y_2,Y_{21})+\alpha_3 I(X_{13},X_{23};Y_3|U,V)\nonumber\\
R_1+R_2\leq&\; \alpha_1I(X_{10};Y_1)+\alpha_2I(X_{20};Y_2,Y_{21})+\alpha_3 I(X_{13},X_{23};Y_3|V)\nonumber\\
R_1+R_2\leq&\; \alpha_1I(X_{10};Y_1,Y_{12})+\alpha_2I(X_{20};Y_2)+\alpha_3 I(X_{13},X_{23};Y_3|U)\nonumber\\
R_1+R_2\leq&\; \alpha_1I(X_{10};Y_1)+\alpha_2I(X_{20};Y_2)+\alpha_3 I(X_{13},X_{23};Y_3)
\end{align}
\emph{for some joint distribution $p(x_{10},u)p(x_{20},v)p(x_{13}|u,v,x_{10})p(x_{23}|u,v,x_{20})P^{\bullet},$ where
$P^{\bullet}$ is the channel given in (\ref{tdchm}), and $\alpha_3=1-\alpha_1-\alpha_2$.}
\begin{proof}
See Appendix C.
\end{proof}
\subsubsection*{Remark $6$} Compared with the achievable region for the PDF scheme in Theorem 1, this outer bound consists of rate pairs $(R_1,R_2)$ satisfying constraints similar to (\ref{th1rr}), except that $I(X_{12};Y_{12})$ and $I(X_{21};Y_{21})$ are replaced by $I(X_{12};Y_1,Y_{12})$ and $I(X_{21};Y_2,Y_{21})$, respectively. These different terms, $I(X_{10};Y_1,Y_{12})$ and $I(X_{20};Y_2,Y_{21})$, resemble the cut-set bound in the first two time slots. Besides, the outer bound has a larger joint input distribution than the achievable region. However, we show in
Section \ref{sec:cap. gau}that for the Gaussian channel, both the achievable region and outer bound can be
maximized over the same input distribution.
\subsection{An Outer Bound Similar to the DF Region}
\subsubsection{Outer Bound Formula}
Following similar steps to the previous outer bound but with small modifications, we derive also in Appendix C another outer bound similar to the DF region as in the following corollary:

\textbf{Corollary $2$.} \emph{An outer bound of the half-duplex MAC-GF consists of the convex hull of all rate pairs $(R_1,R_2)$ satisfying }
\begin{align}\label{outeq}
R_1\leq&\; \alpha_1I(X_{12};Y_1,Y_{12})+\alpha_3 I(X_{13};Y_3|X_{23},S)\nonumber\\
R_2\leq&\; \alpha_2I(X_{21};Y_2,Y_{21})+\alpha_3 I(X_{23};Y_3|X_{13},S)\nonumber\\
R_1+R_2\leq&\; \alpha_1I(X_{12};Y_1,Y_{12})+\alpha_2I(X_{21};Y_2,Y_{21})+\alpha_3 I(X_{13},X_{23};Y_3|S)\nonumber\\
R_1+R_2\leq&\; \alpha_1I(X_{12};Y_1)+\alpha_2I(X_{21};Y_2)+\alpha_3 I(X_{13},X_{23};Y_3)
\end{align}
\emph{for some joint distribution $p(x_{12})p(x_{21})p(s|x_{12},x_{21})p(x_{13}|s,x_{12})p(x_{23}|s,x_{21})P^{\bullet},$ where $P^{\bullet}$ is the channel given in (\ref{tdchm}), and $\alpha_3=1-\alpha_1-\alpha_2$.}

\subsubsection*{Remark $7$} This outer bound looks similar to the achievable region in Theorem 2 for the DF scheme, except for the parts $I(X_{10};Y_1,Y_{12})$ and
$I(X_{20};Y_2,Y_{21})$ where in the achievability, we have $I(X_{10};Y_{12})$ and $I(X_{20};Y_{21})$. As a result of this replacement, the two middle sum rate constraints as in the achievability become redundant.

\subsubsection*{Remark $8$} Similar to the previous outer bound, this outer bound and the DF achievable region for Gaussian channels can be maximized over the same input distribution. For Gaussian channels, as the two achievable rate regions for the PDF and DF schemes are equivalent, the two outer bounds are also equivalent. This equivalence can be proved following similar steps as in Appendix E.

\subsubsection*{Remark $9$} This outer bound is related to the dependence balance outer bound for the full-duplex MAC-GF \cite{dbout} as shown next.
\subsubsection{Relation with Dependence Balance Outer Bound for the Full-Duplex MAC-GF}
In \cite{dbout}, Theorem 4, Tandon and Ulukus derived the dependence balance outer bound for the full-duplex MAC-GF. Applying this outer bound to our half-duplex channel, we can easily derive constraints (\ref{outeq}) as shown in Appendix C. However, for the half-duplex MAC-GF, the dependence balance condition is automatically satisfied as also shown in Appendix C. Hence, the outer bound in Corollary 2
for the half-duplex MAC-GF can be derived from the full-duplex bound but without requiring the dependence balance condition.
\section{Generalization to the $m$-User Half-Duplex MAC-GF}\label{sec:kachout}
\begin{figure}[h]
    \begin{center}
    \includegraphics[width=50mm]{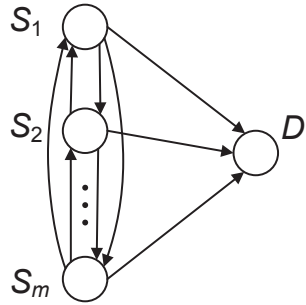}
    \caption{Communication model for the $m$-user half-duplex MAC-GF.} \label{fig:tmuse}
    \end{center}
\vspace*{-7mm}
\end{figure}

We now generalize the results to the $m$-user half-duplex MAC-GF shown in Figure \ref{fig:tmuse}. The m-user half-duplex MAC-GF consists of $m$ input alphabets ${\cal X}_1,{\cal X}_2,\ldots,{\cal X}_m$, $m(m-1)+1$ output alphabets ${\cal Y}$ and ${\cal Y}_{jk}$, for all $j,k\in {1,\ldots,m}$ and $j\neq k$, and $m+1$ transition probabilities $p(y|x_1,\ldots,x_m)$ and $p(y,\boldsymbol{\check{y}}_{jk}|x_j)$  for all $j,k \in {1,\ldots,m}$ and $j\neq k$ where $\boldsymbol{\check{y}}_{jk}=(y_{j1},y_{j2},\ldots,y_{jm}), j\neq k$. Similar to the two user case, we require that no two transition probabilities occur simultaneously in order to meet the half-duplex constraint.
\noindent
\subsection{Achievability}\label{muach}
Following a similar procedure to the two-user case, the transmission is done in independent blocks. Each block is divided into $m+1$ time slots with variable lengths $\alpha_1,\alpha_2,\ldots,\alpha_m,\alpha_{m+1}$ where $\alpha_{m+1}=1-\sum_{i=1}^{m}\alpha_i$. Each user splits its message $w_k$ into two parts $(w_{kk},w_{k,m+1})$. During any of the first $m$ time slots, one of the users sends its cooperative part $w_{kk}$ while the other users decodes it. Then, during the last time slot, each user sends all cooperative parts it decodes during the previous time slots together with its private part $w_{k,m+1}$. Using similar encoding and joint decoding techniques as in the simplified DF scheme in Section \ref{sec: simpsch}, we obtain the following rate region:

\textbf{Theorem $4$.} \emph{An achievable rate region for the m-user half-duplex MAC-GF can be expressed as the convex closure of the rate m-tuples  $(R_1,R_2,\ldots,R_m)$ satisfying}
\begin{align}\label{muhach}
R_{\cal T}\leq&\;\left(\sum_{k\in {\cal T}} \alpha_k \min_{\substack{j\in [1:m],j\neq k}} I(X_{kk};Y_{kj})\right)+\alpha_{m+1} I(\boldsymbol{X}_{m+1}({\cal T});Y_{m+1}|S,\boldsymbol{X}_{m+1}({\cal T}^c))\\
R_{\Omega}\leq&\; \left(\sum_{k\in \Lambda}\alpha_k \min_{\substack{j\in [1:m],j\neq k}} I(X_{kk};Y_{kj})\right)+\left(\sum_{k\in \Lambda^c}\alpha_k I(X_{kk};Y_k)\right)+\alpha_{m+1} I(\boldsymbol{X}_{m+1};Y_{m+1})\nonumber
\end{align}
\noindent \emph{for some joint distribution $P^{\dag}$ given as}
\noindent
\begin{align}
P^{\dag}=\left(\prod_{k=1}^m p(x_{kk})\right)p(s)\left(\prod_{k=1}^m p(x_{k,m+1}|s)\right)
\end{align}
\noindent \emph{and all subsets ${\cal T}\subseteq [1:m]$ and $\Lambda \subseteq [1:m]$. Here, $R_{\cal T}=\sum_{k\in {\cal T}}R_k$, $\boldsymbol{X}_{m+1}=(X_{1,m+1},X_{2,m+1},\ldots,X_{m,m+1})$, $\boldsymbol{X}_{m+1}({\cal T})=(X_{k_1,m+1},X_{k_2,m+1},\ldots)$ for all $\{k_1,k_2,\ldots\}\in {\cal T}$, $\boldsymbol{X}_{m+1}({\cal T}^c)=(X_{l_1,m+1},X_{l_2,m+1},\ldots)$ for all $\{l_1,l_2,\ldots\}\in {\cal T}^c$ and ${\cal T}^c$ is the complement of ${\cal T}$ in the set $[1:m]$. $R_{\Omega}=\sum_{i=1}^m R_{i}$ is the total sum rate and $\Lambda^c$ is the complement of $\Lambda$ in the set $[1:m]$. $Y_{kj}$ is the signal received by user $j$ from user $k$ during the $k^{\text{th}}$ time slot. $Y_k$ is the received signal at the destination during the $k^{\text{th}}$ time slot.}

Furthermore, if each of the inter-user link qualities is better than any link between each user and the destination as given in the following equation:
\begin{align}\label{equ: Kcon}
I(X_{kk};Y_{kj})\geq I(X_{kk};Y_k)\;\text{for all}\;(k,j)\in[1:m], j\neq k
\end{align}
\noindent then, the effective $R_{\Omega}$ becomes only that obtained with empty set $\Lambda=\phi$ and $\Lambda^c=[1:m]$. For example, for 3-user half-duplex MAC-GF satisfying condition (\ref{equ: Kcon}), the achievable rate region is the convex closure of all rate 3-tuples $(R_1,R_2,R_3)$ satisfying
\noindent
\begin{align}
R_1\leq&\; \alpha_1 \min \{I(X_{11},Y_{12}),I(X_{11},Y_{13})\}+\alpha_4 I(X_{14};Y_4|S,X_{24},X_{34})\nonumber\\
R_2\leq&\; \alpha_2 \min \{I(X_{22},Y_{21}),I(X_{22},Y_{23})\}+\alpha_4 I(X_{24};Y_4|S,X_{14},X_{34})\nonumber\\
R_3\leq&\; \alpha_3 \min \{I(X_{33},Y_{31}),I(X_{33},Y_{32})\}+\alpha_4 I(X_{34};Y_4|S,X_{14},X_{24})\nonumber\\
R_1+R_2\leq&\; \alpha_1 \min \{I(X_{11},Y_{12}),I(X_{11},Y_{13})\} + \alpha_2 \min \{I(X_{22},Y_{21}),I(X_{22},Y_{23})\}+\alpha_4 I(X_{14},X_{24};Y_4|S,X_{34})\nonumber\\
R_1+R_3\leq&\; \alpha_1 \min \{I(X_{11},Y_{12}),I(X_{11},Y_{13})\} + \alpha_3 \min \{I(X_{33},Y_{31}),I(X_{33},Y_{32})\}+\alpha_4 I(X_{14},X_{34};Y_4|S,X_{24})\nonumber\\
R_2+R_3\leq&\; \alpha_2 \min \{I(X_{22},Y_{21}),I(X_{22},Y_{23})\} + \alpha_3 \min \{I(X_{33},Y_{31}),I(X_{33},Y_{32})\}+\alpha_4 I(X_{14},X_{34};Y_4|S,X_{24})\nonumber\\
R_1+R_2+R_3\leq&\; \alpha_1 \min \{I(X_{11},Y_{12}),I(X_{11},Y_{13})\} + \alpha_2 \min \{I(X_{22},Y_{21}),I(X_{22},Y_{23})\}\nonumber\\
+&\;\alpha_3 \min \{I(X_{33},Y_{31}),I(X_{33},Y_{32})\}+\alpha_4 I(X_{14},X_{24},X_{34};Y_4|S)\nonumber\\
R_1+R_2+R_3\leq&\; \alpha_1 I(X_{11},Y_1) + \alpha_2I(X_{22},Y_2)+\alpha_3 I(X_{33},Y_3)+\alpha_4 I(X_{14},X_{24},X_{34};Y_4)
\end{align}
\noindent for some joint distribution $p(x_{11})p(x_{22})p(x_{33})p(s)p(x_{14}|s)p(x_{24}|s)p(x_{34}|s)$. 
\subsection{Outer Bound}
An outer bound for the $m$-user half-duplex MAC-GF can also be derived in a similar way to that of the two-user case in Section \ref{sec:out. cap}. Using Fano's and data processing inequalities among multiple time slots as given in Appendix C for the two-user case, we can obtain an outer bound as in the following Theorem:

\textbf{Theorem $5$.} \emph{An outer bound for the $m$-user half-duplex MAC-GF can be expressed as the convex closure of all rate m-tuples  $(R_1,R_2,\ldots,R_m)$ satisfying rate constraints obtained from (\ref{muhach}) by replacing}
\begin{itemize}
\item
$\sum_{k\in {\cal T}} \alpha_k \min_{\substack{j\in [1:m],j\neq k}} I(X_{kk};Y_{kj})$ by $\sum_{k\in {\cal T}} \alpha_k I(X_{kk};Y_k,\boldsymbol{Y\!\!}_{kj})$ and
\item
$\sum_{k\in \Lambda}\alpha_k \min_{\substack{j\in [1:m],j\neq k}} I(X_{kk};Y_{kj})$ by $\sum_{k\in \Lambda}\alpha_k I(X_{kk};Y_k,\boldsymbol{Y\!\!}_{kj})$
\end{itemize}
\noindent \emph{where $\boldsymbol{Y\!\!}_{kj}=(Y_{k,j_1},Y_{k,j_2}),\ldots$ for all $\{j_1,j_2,\ldots\}\in [1:m]$ and $j\neq k$. Furthermore, the joint probability distribution factors as}
\begin{align}
P^{\ddagger}=\left(\prod_{k=1}^m p(x_{kk})\right)p(s|x_{11},x_{22},\ldots,x_{mm})\left(\prod_{k=1}^m p(x_{km+1}|s,x_{kk})\right)\breve{P}
\end{align}
\noindent\emph{where $\breve{P}$ is the channel given as}

\noindent
\begin{align}
\breve{P}=&\;\left(\sum_{k=1}^m p(y_k,\check{y}_{kj}|x_{kk})\left(u\left(\tau-(\sum_{l=1}^{k-1}\alpha_l)n\right)-u\left(\tau-(\sum_{l=1}^k\alpha_l)n\right)\right)\right)\nonumber\\
&+\;p(y_{m+1}|\boldsymbol{x}_{m+1})\left(u\left(\tau-(\sum_{i=1}^{m}\alpha_i)n\right)-u(\tau-n)\right)
\end{align}
\noindent \emph{where $\boldsymbol{x}_{m+1}=(x_{1,m+1},x_{2,m+1},\ldots,x_{m,m+1})$.}

Although the outer bound for the two-user case is tight especially for the Gaussian channel, it becomes looser as the number of users increases. Even for the three-user case, the outer bound is not tight in general. This is mainly because during the first $m$ time slots, the outer is a cut set bound while in the achievability, it is the  rate achieved by the user with minimum link quality.
Hence, the optimal coding scheme and the tightest outer bound are still an open problem for the $m$-user half-duplex
MAC-GF.
\section{Gaussian Channels}\label{sec:cap. gau}
\subsection{The Half-Duplex Gaussian MAC-GF Model}
\begin{figure}[h]
    \begin{center}
    \includegraphics[width=110mm]{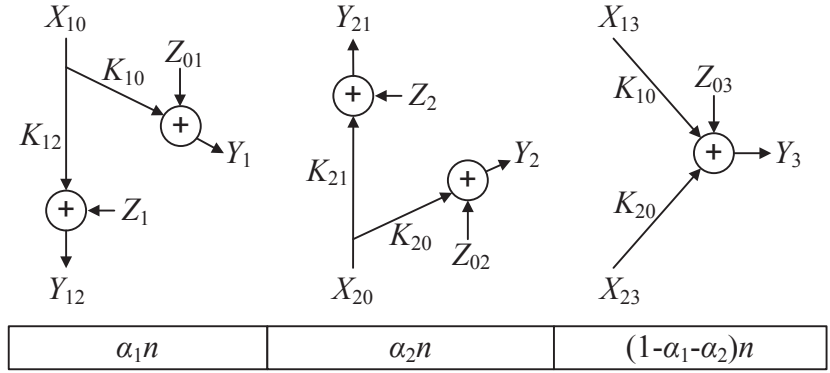}
    \caption{Channel model for the Gaussian MAC-GF.} \label{fig:Gausmod}
    \end{center}
\vspace*{-7mm}
\end{figure}

The discrete-time channel model for the Gaussian half-duplex MAC-GF, as shown in Fig. \ref{fig:Gausmod}, can be expressed as
\noindent
\begin{align}\label{Gchm}
Y_{12}&=K_{12}X_{10}+Z_1\nonumber\\
Y_{21}&=K_{21}X_{20}+Z_2\nonumber\\
Y_{1}&=K_{10}X_{10}+Z_{01}\nonumber\\
Y_{2}&=K_{20}X_{20}+Z_{02}\nonumber\\
Y_{3}&=K_{10}X_{13}+K_{20}X_{23}+Z_{03}
\end{align}
\noindent where $K_{12}$ and $K_{21}$ are the inter-user link coefficients; $K_{10},$ and $K_{20}$ are the link coefficients between each user and the destination; the independent AWGNs are $Z_1\sim N(0,N), Z_2\sim N(0,N),$ and $Z_{0i}\sim N(0,N),\; i=1,2,3$; $X_{10}$ and $X_{13}$ are the first user's transmitted signals during the $1^\text{st}$ and $3^\text{rd}$ time slots, respectively, similarly,  $X_{20}$ and $X_{23}$ are the second user's transmitted signals during the $2^\text{nd}$ and $3^\text{rd}$ time slots.
\noindent
\subsection{Partial Decode-forward scheme: Joint decoding vs. Separate decoding}\label{Gach}
Using the PDF scheme, we show in Appendix D that jointly Gaussian input signals are optimal. Specifically, the first user can construct its transmitted signals as
\noindent
\begin{align*}
X_{10}&=\sqrt{P_{10}}\check{X}_{10}(w_{10})+\sqrt{P_{U}}U(w_{12})\\
X_{13}&=\sqrt{P_{13}}\check{X}_{13}(w_{13})+\sqrt{c_2P_{U}}U(w_{12})+\sqrt{c_3P_{V}}V(w_{21})
\end{align*}
\noindent and the second user constructs its transmitted signals as
\noindent
\begin{align*}
X_{20}&=\sqrt{P_{20}}\check{X}_{20}(w_{20})+\sqrt{P_{V}}V(w_{21})\\
X_{23}&=\sqrt{P_{23}}\check{X}_{23}(w_{23})+\sqrt{d_2P_{V}}V(w_{21})+\sqrt{d_3P_{U}}U(w_{12}),
\end{align*}
\noindent where $\check{X}_{10},\check{X}_{20},\check{X}_{13},\check{X}_{23},U,$ and $V$ are independent and identically distributed according to $N(0,1)$.

The power constraints for the two users are given as
\noindent
\begin{align}\label{powc}
\alpha_1(P_{10}+P_{U})+\alpha_3(P_{13}+c_2P_{U}+c_3P_{V})&=P_1\nonumber\\
\alpha_2(P_{20}+P_{V})+\alpha_3(P_{23}+d_3P_{U}+d_2P_{V})&=P_2
\end{align}
\noindent where $(c_2, c_3)$ are constant factors specifying the relative amount of power, compared to $P_{U}$ and $P_{V}$, used by the first user to transmit the cooperative information $(w_{12}, w_{21})$ during the $3^\text{rd}$ time slot. The same holds for $(d_2, d_3)$.


The achievable rate region for the PDF scheme over Gaussian channels with each of the decoding techniques in Section \ref{sec:achrr} can be derived as follows.\\

\subsubsection{PDF with Joint Decoding}\label{gpar}
The achievable rate region for this decoding technique is given in (\ref{th1rr}) of Theorem 1. Applying to the Gaussian channel, we obtain the following corollary:

\textbf{Corollary $3$.} \emph{The achievable rate region for the Gaussian half-duplex MAC-GF under the PDF scheme with full decoding at each user and joint decoding at the destination
is the union of all rate pairs $(R_1,R_2)$ satisfying }
\begin{align}\label{th1Grr}
R_1\leq&\; \alpha_1C\left(\frac{K_{12}^2\left(P_{U}+P_{10}\right)}{N}\right)+\alpha_3C\left(\frac{K_{10}^2P_{13}}{N}\right)\nonumber\\
R_2\leq&\; \alpha_2C\left(\frac{K_{21}^2\left(P_{V}+P_{20}\right)}{N}\right)+\alpha_3C\left(\frac{K_{20}^2P_{23}}{N}\right)\nonumber\\
R_1+R_2\leq&\; \alpha_1C\left(\frac{K_{12}^2\left(P_{U}+P_{10}\right)}{N}\right)+\alpha_2C\left(\frac{K_{21}^2\left(P_{V}+P_{20}\right)}{N}\right)
+\alpha_3C\left(\frac{K_{10}^2P_{13}+K_{20}^2P_{23}}{N}\right)\nonumber\\
R_1+R_2\leq&\; \alpha_1C\left(\frac{K_{10}^2\left(P_{U}+P_{10}\right)}{N}\right)+\alpha_2C\left(\frac{K_{21}^2\left(P_{V}+P_{20}\right)}{N}\right)\nonumber\\
&+\alpha_3C\left(\frac{K_{10}^2(P_{13}+c_2P_{U})+K_{20}^2(P_{23}+d_3P_{U})+2K_{10}K_{20}\sqrt{c_2d_3}P_{U}}{N}\right)\nonumber\\
R_1+R_2\leq&\; \alpha_1C\left(\frac{K_{12}^2\left(P_{U}+P_{10}\right)}{N}\right)+\alpha_2C\left(\frac{K_{20}^2\left(P_{V}+P_{20}\right)}{N}\right)\nonumber\\
&+\alpha_3C\left(\frac{K_{10}^2(P_{13}+c_3P_{V})+K_{20}^2(P_{23}+d_2P_{V})+2K_{10}K_{20}\sqrt{d_2c_3}P_{V}}{N}\right)\nonumber\\
R_1+R_2\leq&\; \alpha_1C\left(\frac{K_{10}^2\left(P_{U}+P_{10}\right)}{N}\right)+\alpha_2C\left(\frac{K_{20}^2\left(P_{V}+P_{20}\right)}{N}\right)\nonumber\\
&+\alpha_3C\left(\frac{K_{10}^2P_{13}+K_{20}^2P_{23}+P_U(K_{10}\sqrt{c_2}+K_{20}\sqrt{d_3})^2+P_V(K_{10}\sqrt{c_3}+K_{20}\sqrt{d_2})^2}{N}\right)
\end{align}
\noindent \emph{for some $\alpha_1,\alpha_2\geq 0,$ $\alpha_1+\alpha_2\leq 1$ and power allocation $(P_U,P_V,P_{10},P_{20},P_{13},P_{23})$ satisfying the power constraint in
(\ref{powc}), where $C(x)=0.5\log(1+x)$.}
\subsubsection*{Remark $10$: Achievable rate region with partial decoding at each user}
The achievable rate region for the Gaussian MAC-GF with partial decoding at each user is similar to (\ref{th1Grr}) but replacing
\begin{itemize}
\item
$C\left(\frac{K_{12}^2\left(P_{U}+P_{10}\right)}{N}\right)$ by $C\left(\frac{K_{12}^2P_U}{K_{12}^2P_{10}+N}\right)+C\left(\frac{K_{10}^2P_{10}}{N}\right)$ and
\item
$C\left(\frac{K_{21}^2\left(P_{V}+P_{20}\right)}{N}\right)$ by $C\left(\frac{K_{21}^2P_V}{K_{12}^2P_{10}+N}\right)+C\left(\frac{K_{20}^2P_{20}}{N}\right)$
\end{itemize}
\noindent As explained in Section \ref{secpdu}, in order to compare between the rate regions with full and partial decoding, we only need to compare between  $C\left(\frac{K_{10}^2P_{10}}{N}\right)$ and $C\left(\frac{K_{12}^2P_{10}}{N}\right)$, which is equivalent to comparing between $K_{10}$ and $K_{12}$. Hence we can see that for any input distribution, if $K_{12}>K_{10}$ and $K_{21}>K_{20}$, then the rate region with full decoding is bigger. However,
following the optimization in \cite{haivu2}, the two rate regions can be shown to be equivalent if each user
transmits with an optimal power allocation which has $P_{10}=P_{20}=0$.
\subsubsection{PDF with Separate Decoding}
For the Gaussian channel, the achievable rate region of the PDF scheme with separate decoding can be obtained directly from (\ref{co2rr}) as follows.

\textbf{Corollary $4$.} \emph{The achievable rate region for the Gaussian half-duplex MAC-GF under the PDF scheme with \textbf{separate} decoding at the destination is the union of all rate pairs $(R_1,R_2)$ satisfying}
\noindent
\begin{align}\label{co2Grr}
R_1\leq&\; \alpha_1C\left(\frac{K_{12}^2\left(P_{U}+P_{10}\right)}{N}\right)+\alpha_3C\left(\frac{K_{10}^2P_{13}}{N}\right)\nonumber\\
R_2\leq&\; \alpha_2C\left(\frac{K_{21}^2\left(P_{V}+P_{20}\right)}{N}\right)+\alpha_3C\left(\frac{K_{20}^2P_{23}}{N}\right)\nonumber\\
R_1+R_2\leq&\; \alpha_1C\left(\frac{K_{12}^2\left(P_{U}+P_{10}\right)}{N}\right)+\alpha_2C\left(\frac{K_{21}^2\left(P_{V}+P_{20}\right)}{N}\right)+\alpha_3C\left(\frac{K_{10}^2P_{13}+K_{20}^2P_{23}}{N}\right)\nonumber\\
R_1+R_2\leq&\; \alpha_1 \text{min}\left\{C\left(\frac{K_{12}^2P_{10}}{N_1}\right),\;C\left(\frac{K_{10}^2P_{10}}{N}\right)\right\}+\alpha_2C\left(\frac{K_{21}^2\left(P_{V}+P_{20}\right)}{N}\right)\nonumber\\
&+\alpha_3 C\left(\frac{K_{10}^2(P_{13}+c_2P_{U})+K_{20}^2(P_{23}+d_3P_{U})+2K_{10}K_{20}\sqrt{c_2d_3}P_{U}}{N}\right)
\nonumber\\
R_1+R_2\leq&\; \alpha_1C\left(\frac{K_{12}^2\left(P_{U}+P_{10}\right)}{N}\right)+\alpha_2\text{min}\left\{C\left(\frac{K_{21}^2P_{20}}{N}\right),\;C\left(\frac{K_{20}^2P_{20}}{N}\right)\right\}\nonumber\\
&+\alpha_3 C\left(\frac{K_{10}^2(P_{13}+c_3P_{V})+K_{20}^2(P_{23}+d_2P_{V})+2K_{10}K_{20}\sqrt{d_2c_3}P_{V}}{N}\right)
\nonumber\\
R_1+R_2\leq&\;\alpha_1\text{min}\left\{C\left(\frac{K_{12}^2P_{1}}{N}\right),\;C\left(\frac{K_{10}^2P_{10}}{N}\right)\right\}+\alpha_2\text{min}\left\{C\left(\frac{K_{21}^2P_{20}}{N}\right),\;C\left(\frac{K_{20}^2P_{20}}{N_0}\right)\right\}\nonumber\\
&+\alpha_3 C\left(\frac{K_{10}^2P_{13}+K_{20}^2P_{23}+P_U(K_{10}\sqrt{c_2}+K_{20}\sqrt{d_3})^2+P_V(K_{10}\sqrt{c_3}+K_{20}\sqrt{d_2})^2}{N}\right)
\end{align}
\noindent \emph{for some $\alpha_1,\alpha_2\geq 0$ and $\alpha_1+\alpha_2<1$, given the power constraint in (\ref{powc}).}
\subsubsection*{Remark $11$: Comparison between joint and separate decoding}
From (\ref{th1Grr}) and (\ref{co2Grr}), we can see that joint decoding leads to a strictly larger rate region than separate decoding because
\begin{align*}
C\left(\frac{K_{10}^2\left(P_{U}+P_{10}\right)}{N}\right)\geq &\text{min}\left\{C\left(\frac{K_{12}^2P_{10}}{N}\right),\;C\left(\frac{K_{10}^2P_{10}}{N}\right)\right\}\\
C\left(\frac{K_{20}^2\left(P_{V}+P_{20}\right)}{N}\right)\geq &\text{min}\left\{C\left(\frac{K_{21}^2P_{20}}{N}\right),\;C\left(\frac{K_{20}^2P_{20}}{N}\right)\right\}
\end{align*}

\begin{figure}[t]
    \begin{center}
    \includegraphics[width=0.6\textwidth, height=80mm]{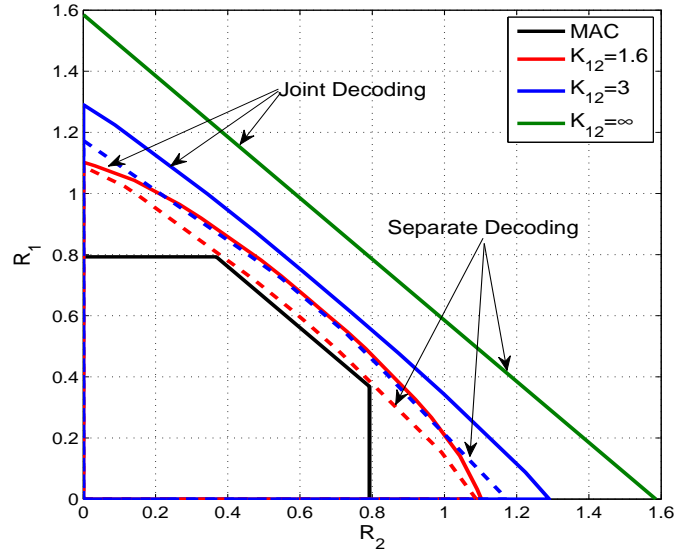}
    \caption{Achievable rate regions for the half-duplex MAC-GF using the PDF scheme with joint and separate decoding for $K_{10}=K_{20}=1$ and $K_{12}=K_{21}$.} \label{fig:ratejs}
    \end{center}
\vspace*{-8mm}
\end{figure}

Fig. \ref{fig:ratejs} compares the achievable rate regions of the PDF scheme with different decoding techniques and the classical MAC. These results are obtained for the symmetric Gaussian channel where
$N=1, P_1=P_2=2$, $K_{10}=K_{20}=1,$ $K_{12}=K_{21}$ and different values of $K_{12}$ and by using the
optimal power allocations and time durations analyzed in \cite{haivu2}. Results show that the PDF scheme with either joint or separate decoding at the destination has a larger rate
region than the MAC, and the rate region enlarges as $K_{12}$ increases. The results also show the advantage of joint decoding over separate decoding. Separate decoding is strictly suboptimal compared to joint decoding and only approaches the performance of joint decoding as $K_{12}\rightarrow \infty$.
\noindent

\subsection{Decode-forward scheme}
The simplified DF scheme in Section \ref{sec: simpsch} can be applied to a Gaussian channel with the first and second users constructing their transmitted signals as
\noindent
\begin{align*}
X_{12}&=\sqrt{P_{12}}\check{X}_{12}(w_{12})\\
X_{13}&=\sqrt{P_{13}}\check{X}_{13}(w_{13})+\sqrt{P_{S_1}}S(w_{12},w_{21})\\
X_{21}&=\sqrt{P_{21}}\check{X}_{21}(w_{12})\\
X_{23}&=\sqrt{P_{23}}\check{X}_{23}(w_{13})+\sqrt{P_{S_2}}S(w_{21},w_{21})
\end{align*}
\noindent where $\check{X}_{12},\check{X}_{21},\check{X}_{13},\check{X}_{23}$ and $S$ are independent and identically distributed according to $N(0,1)$.
The power constraints for both users are now given as
\noindent
\begin{align}\label{powcsch2}
\alpha_1P_{12}+\alpha_3(P_{13}+P_{S_1})&=P_1\nonumber\\
\alpha_2P_{21}+\alpha_3(P_{23}+P_{S_2})&=P_2
\end{align}
The achievable rate region can be derived as follows.

\textbf{Corollary $5$.} \emph{The achievable rate region for the Gaussian half-duplex MAC-GF under the DF scheme,  denoted as ${\cal R}(K_{12}^2,K_{21}^2)$, is given as}
\noindent
\begin{align}
R_1\leq&\; \alpha_1C\left(\frac{K_{12}^2P_{12}}{N}\right)+\alpha_3C\left(\frac{K_{10}^2P_{13}}{N}\right)\nonumber\\
R_2\leq&\; \alpha_2C\left(\frac{K_{21}^2P_{21}}{N}\right)+\alpha_3C\left(\frac{K_{20}^2P_{23}}{N}\right)\nonumber\\
R_1+R_2\leq&\; \alpha_1C\left(\frac{K_{12}^2P_{12}}{N}\right)+\alpha_2C\left(\frac{K_{21}^2P_{21}}{N}\right)+\alpha_3C\left(\frac{K_{10}^2P_{13}+K_{20}^2P_{23}}{N}\right)\nonumber
\end{align}
\begin{align}\label{th1Grrsimp}
R_1+R_2\leq&\; \alpha_1C\left(\frac{K_{10}^2P_{12}}{N}\right)+\alpha_2C\left(\frac{K_{21}^2P_{21}}{N}\right)\nonumber\\
&+\alpha_3C\left(\frac{K_{10}^2(P_{13}+P_{S_1})+K_{20}^2(P_{23}+P_{S_2})+2K_{10}K_{20}\sqrt{P_{S_1}P_{S_2}}}{N}\right)\nonumber\\
R_1+R_2\leq&\; \alpha_1C\left(\frac{K_{12}^2P_{12}}{N}\right)+\alpha_2C\left(\frac{K_{20}^2P_{21}}{N}\right)\nonumber\\
&+\alpha_3C\left(\frac{K_{10}^2(P_{13}+P_{S_1})+K_{20}^2(P_{23}+P_{S_2})+2K_{10}K_{20}\sqrt{P_{S_1}P_{S_2}}}{N}\right)\nonumber\\
R_1+R_2\leq&\; \alpha_1C\left(\frac{K_{10}^2P_{12}}{N}\right)+\alpha_2C\left(\frac{K_{20}^2P_{21}}{N}\right)\nonumber\\
&+\alpha_3C\left(\frac{K_{10}^2(P_{13}+P_{S_1})+K_{20}^2(P_{23}+P_{S_2})+2K_{10}K_{20}\sqrt{P_{S_1}P_{S_2}}}{N}\right)
\end{align}
\noindent \emph{for some some $\alpha_1,\alpha_2\geq 0,$ $\alpha_1+\alpha_2<1$ and power allocation $(P_{12},P_{21},P_{13},P_{23},P_{S_1},P_{S_2})$ satisfying the power constraint given in (\ref{powcsch2}).}

Although the rate region of the DF scheme in (\ref{th1Grrsimp}) appears to be larger than that of the PDF scheme in (\ref{th1Grr}), they are equivalent as shown in Appendix E. We state this equivalence in the following corollary:

\textbf{Corollary $6$.} \emph{The two achievable rate regions for the Gaussian half-duplex MAC-GF under the PDF scheme and
the DF scheme are equivalent.}
\begin{proof}
See Appendix E.
\end{proof}
\subsection{Outer Bound}
Similar to the equivalence between the two achievable regions, the two outer bounds in Theorem $3$ and Corollary $2$ are also equivalent for the Gaussian channel. This can be shown using the same procedure as in Appendix E. Therefore, in this section, we only focus on the outer in (\ref{outeq}) bound with rate constraints similar to the DF scheme.

\textbf{Corollary $7$.} \emph{An outer bound for the Gaussian half-duplex MAC-GF is ${\cal R}(K_{12}^2+K_{10}^2,K_{21}^2+K_{20}^2)$, which consists of all rate pairs $(R_1,R_2)$ satisfying (\ref{th1Grrsimp}) but replacing $K_{12}^2$ by $K_{12}^2+K_{10}^2$ and $K_{21}^2$ by $K_{21}^2+K_{20}^2$.}
\begin{proof}
As stated in Corollary $2$, the outer bound is maximized over the joint distribution $p(x_{12})p(x_{21})p(x_{13}|s,x_{12})$ $p(x_{23}|s,x_{21})P^{\bullet}$. The difference between this distribution and that of the achievable region in Corollary $5$ is that $X_{13}$ and $X_{12}$ ($X_{23}$ and $X_{21}$) are correlated. Let $\sqrt{L_{12}}$  $(\sqrt{L_{21}})$ be the correlation between $X_{13}$ and $X_{12}$ ($X_{23}$ and $X_{21}$). Then, from (\ref{outeq}) we obtain the same expressions as in (\ref{th1Grrsimp}) but only replacing
\begin{itemize}
\item
$P_{13}$ by $\acute{P}_{13}+L_{12}$ and $P_{23}$ by $\acute{P}_{23}+L_{21}$;
\item
$K_{12}^2$ by $K_{12}^2+K_{10}^2$ and $K_{21}^2$ by $K_{21}^2+K_{20}^2$
\end{itemize}
Since the power constraints are the same for the achievable region and outer bound, it is possible to set
$P_{13}=\acute{P}_{13}+L_{12}$ and $P_{23}=\acute{P}_{23}+L_{21}$ in the outer bound,  thus reducing the input distribution to that of the achievable region.  Therefore, the only remaining difference is in the channel gain.
\end{proof}
\subsubsection*{Remark $12$} The tightness of the outer bound is determined by the ratios $\frac{K_{12}^2}{K_{10}^2}$ and
$\frac{K_{21}^2}{K_{20}^2}$. The outer bound becomes tighter as these two ratios increase since then $K_{12}^2\rightarrow K_{12}^2+K_{10}^2$ and $K_{21}^2\rightarrow K_{21}^2+K_{20}^2$.
Therefore, the bound becomes increasingly tight as the inter-user link qualities increase.

\begin{figure}[t]
    \begin{center}
    \includegraphics[width=0.6\textwidth, height=80mm]{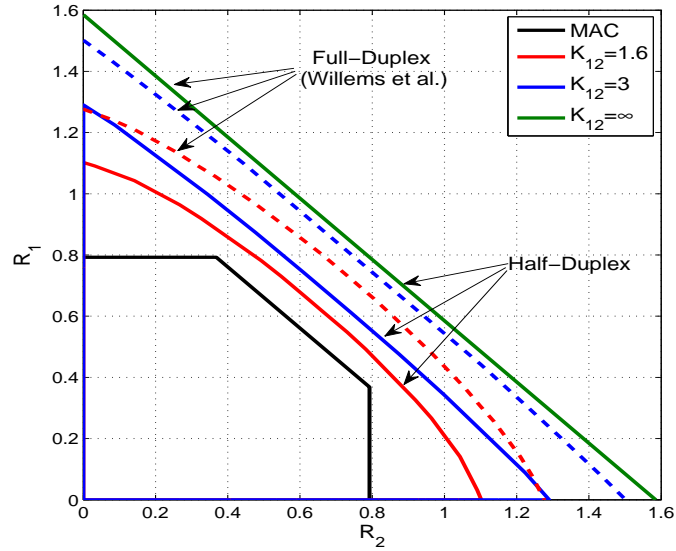}
    \caption{Achievable rate regions for the half-duplex MAC-GF compared with full-duplex and the classical MAC $(K_{10}=K_{20}=1, K_{12}=K_{21})$.} \label{fig:ratefh}
    \end{center}
\vspace*{-10mm}
\end{figure}

Fig. \ref{fig:ratefh} compares between the DF scheme with joint decoding and the full-duplex scheme in \cite{fcc2005fof,cognitiveRT}. As expected, the half-duplex scheme has a smaller
rate region than the full-duplex scheme. The two regions become closer to each other as
$K_{12}$ increases. However, in the full-duplex scheme, each user transmits and receives in two different frequency bands \cite{cognitiveRT} which doubles the required bandwidth.

\begin{figure}[t]
    \begin{center}
    \includegraphics[width=0.6\textwidth, height=80mm]{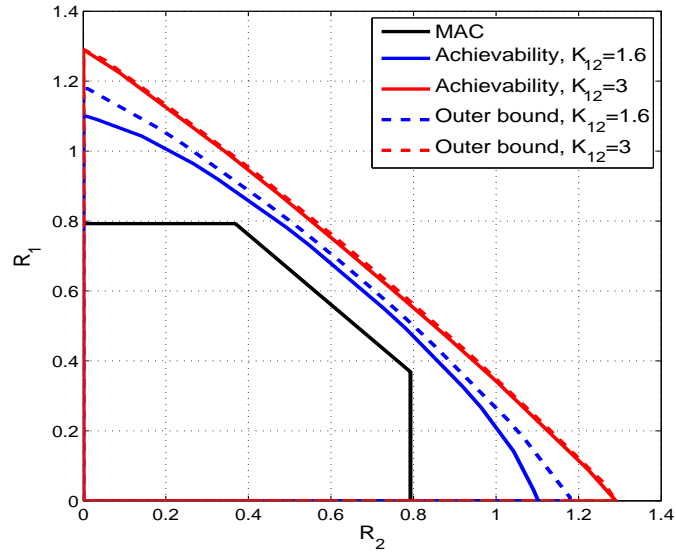}
    \caption{Achievable rate regions and outer bounds for the symmetric half-duplex MAC-GF with $K_{10}=K_{20}=1$ and  $K_{12}=K_{21}$.} \label{fig:raout}
    \end{center}
\vspace*{-8mm}
\end{figure}

\begin{figure}[t]
    \begin{center}
    \includegraphics[width=0.6\textwidth, height=80mm]{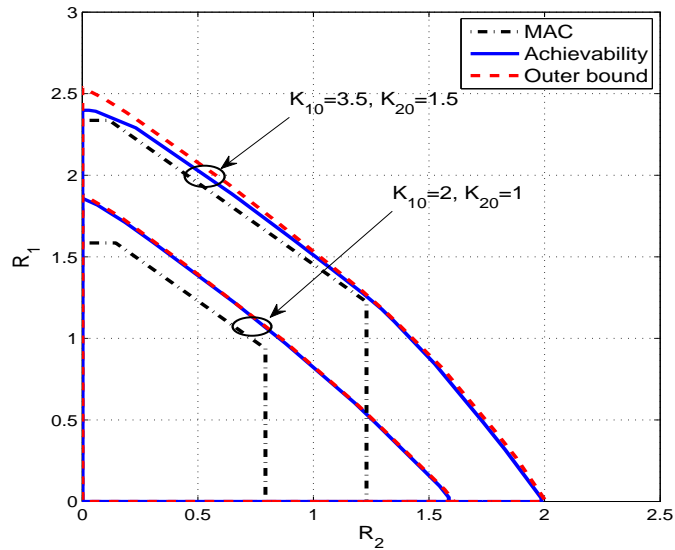}
    \caption{Achievable rate regions and outer bounds for asymmetric half-duplex MAC-GF with $K_{12}=K_{21}=4$.}
    \label{fig:asyro}
    \end{center}
\vspace*{-10mm}
\end{figure}

Figures \ref{fig:raout} and \ref{fig:asyro} compare between the achievable rate region for the DF scheme and the outer bound for both cases of symmetric and asymmetric half-duplex MAC-GF. In Fig. \ref{fig:raout}, results are plotted for the symmetric case with different values of $K_{12}$ while $K_{10}=1$. In Fig. \ref{fig:asyro},  results are plotted for the asymmetric case with different values of $K_{10}$ and $K_{20}$ while $K_{12}=K_{21}=4$. As discussed in Remark $12$,
these results show that the achievable rate region becomes closer to the outer bound as
the ratios $\frac{K_{12}^2}{K_{10}^2}$ and $\frac{K_{21}^2}{K_{20}^2}$ increase.
\noindent
\noindent

\subsection{Capacity for the Physically Degraded Gaussian Channel}
For the physically degraded Gaussian channel where $I(X_{12};Y_1|Y_{12})=I(X_{21};Y_2|Y_{21})=0$ for all $p(x_{12})p(x_{21})$, the achievable region becomes the capacity. The realization of the degraded Gaussian channel applies to a channel having correlated noise with a specific correlation factor. Without loss of generality, assume that the AWGN noises in (\ref{Gchm}) are identically distributed with zero mean and unit variance. Let $Z_1$ and $Z_{01}$ be correlated with correlation factor $\rho_1$. Similarly, let $Z_2$ and $Z_{02}$ be correlated with correlation factor $\rho_2$. While these correlations do not affect the achievability in Corollary 4, they alter the outer bound in Theorem $3$ as follows.
\begin{align*}
I(X_{12};Y_1,Y_{12})=&\;C\left(\frac{\left(K_{12}^2+K_{10}^2-2K_{10}K_{12}\rho_1\right)P_{12}}{1-\rho_1^2}\right)\\
I(X_{21};Y_2,Y_{21})=&\;C\left(\frac{\left(K_{21}^2+K_{20}^2-2K_{20}K_{21}\rho_2\right)P_{21}}{1-\rho_2^2}\right)
\end{align*}
Comparing these equations with the achievable counterparts:
\begin{align*}
I(X_{12};Y_{12})=&\;C\left(K_{12}^2P_{12}\right)\\
I(X_{21};Y_{21})=&\;C\left(K_{21}^2P_{21}\right),
\end{align*}
\noindent we can easily show that they are equal if $\rho_1=\frac{K_{10}}{K_{12}},\;\rho_2=\frac{K_{20}}{K_{21}}$ and $K_{12}>K_{10},\;K_{21}>K_{20}$. Therefore, we have the following result:

\textbf{Theorem $6$.} \emph{The capacity region for the physically degraded Gaussian half-duplex MAC-GF can be achieved using either the PDF or DF coding schemes if $(Z_1,Z_{01})$ and $(Z_2,Z_{02})$ are correlated with correlation factors $\frac{K_{10}}{K_{12}}$ and $\frac{K_{20}}{K_{21}}$, respectively, provided that $K_{12}>K_{10}$ and $K_{21}>K_{20}$.}



\section{Conclusion}\label{sec:conclusion}
In this paper, we have analyzed achievable regions and outer bounds for the half-duplex MAC-GF. We propose two coding schemes (PDF and DF) based on rate splitting and superposition encoding. In the PDF scheme, each user splits its message into $3$ parts and the codewords in the $3^{\text{rd}}$ time slot are superimposed on those of the first two. In the DF scheme, each user splits its message into $2$ parts and has independent codewords at each time slot. For the PDF scheme, we analyze the advantage of joint decoding over separate decoding at the destination. The DF scheme, however, is simpler and leads to the same region for the Gaussian channel which models many practical channels. Therefore, DF is preferred for practical implementation and for further analysis of the optimal power allocation and time duration.

We also derive two outer bounds with rate constraints similar to each of the two achievable regions using standard Fano's and data processing inequalities. We show the equivalence between these two outer bounds for the Gaussian channel. We also show that one of the outer bounds can be derived directly from the dependence balance outer bound of the full-duplex channel but without explicit dependence balance constraint. With straightforward generalization, we extended out results to the $m$-user case.

Lastly, we have presented numerical examples for the Gaussian channel
that compare between the proposed schemes, the classical MAC, the full-duplex MAC-GF, and the outer bound. These results also show the advantage of joint decoding compared to separate decoding. When the inter-user link is better than the link between each user and the destination, results show that cooperation
improves the rate region over the classical MAC even with half-duplex constraint. Moreover, the rate region approaches the outer bound as the inter-user link quality increases.

\appendices
\section{Error Analysis of the PDF Scheme with Joint Decoding at the Destination}

Because of the symmetry of the random code generation, the conditional error probability does not depend on which message vector was sent. Hence, without loss of generality, we assume that the message vector $w_{12}=w_{21}=w_{10}=w_{20}=w_{13}=w_{23}=1$ was sent. Then, the error events at the destination are as follows.
\begin{align}
E_1:=&\{(u^{\alpha_1 n}(1), x_{10}^{\alpha_1 n}(1,1),\boldsymbol{Y\!}_1)\not\in A_\epsilon ^{\alpha_1 n}\};\nonumber\\
E_2:=&\{(u^{\alpha_1 n}(1), x_{10}^{\alpha_1 n}(w_{10},1),\boldsymbol{Y\!}_1)\in A_\epsilon ^{\alpha_1 n} \;\text{for some}\;w_{10}\neq1\};\nonumber\\
E_3:=&\{(v^{\alpha_2 n}(1), x_{20}^{\alpha_2 n}(1,1),\boldsymbol{Y\!}_2)\not\in A_\epsilon ^{\alpha_2 n}\};\nonumber\\
E_4:=&\{(v^{\alpha_2 n}(1), x_{20}^{\alpha_2 n}(w_{20},1),\boldsymbol{Y\!}_2)\in A_\epsilon ^{\alpha_2 n} \;\text{for some}\;w_{20}\neq1\};\nonumber\\
E_5:=&\{(u^{\alpha_3 n}(1),v^{\alpha_3 n}(1),x_{13}^{\alpha_3n}(1,1,1),x_{23}^{\alpha_3 n}(1,1,1), \boldsymbol{Y\!}_3)\not\in A_\epsilon ^{\alpha_3 n}\};\nonumber\\
E_6:=&\{(u^{\alpha_3 n}(1),v^{\alpha_3 n}(1),x_{13}^{\alpha_3 n}(w_{13},1,1),x_{23}^{\alpha_3 n}(1,1,1), \boldsymbol{Y\!}_3)\in A_\epsilon ^{\alpha_3 n}\;\text{for some}\;w_{13}\neq1\};\nonumber\\
E_7:=&\{(u^{\alpha_3 n}(1),v^{\alpha_3 n}(1),x_{13}^{\alpha_3 n}(1,1,1),x_{23}^{\alpha_3 n}(w_{23},1,1), \boldsymbol{Y\!}_3)\in A_\epsilon ^{\alpha_3 n}\;\text{for some}\;w_{23}\neq1\};\nonumber\\
E_8:=&\{(u^{\alpha_3 n}(1),v^{\alpha_3 n}(1),x_{13}^{\alpha_3 n}(w_{13},1,1),x_{23}^{\alpha_3 n}(w_{23},1,1), \boldsymbol{Y\!}_3)\in A_\epsilon ^{\alpha_3 n}\;\text{for some}\;
(w_{13}\neq1, w_{23}\neq1)\};\nonumber\\
E_9:=&\{(u^{\alpha_1n}(1),x_{10}^{\alpha_1 n}(1,1),\boldsymbol{Y\!}_1)\not\in A_\epsilon ^{\alpha_1n},\;\text{and}\;(v^{\alpha_2n}(1),x_{20}^{\alpha_2 n}(1,1),\boldsymbol{Y\!}_2)\not\in A_\epsilon ^{\alpha_2n}, \text{and}\nonumber\\
&(u^{\alpha_3 n}(1),v^{\alpha_3 n}(1),x_{13}^{\alpha_3 n}(1,1,1),x_{23}^{\alpha_3 n}(1,1,1),\boldsymbol{Y\!}_3)\not\in A_\epsilon ^{\alpha_3 n}\};\nonumber\\
E_{10}:=&\{(u^{\alpha_1n}(w_{12}),x_{10}^{\alpha_1 n}(w_{10},w_{12}),\boldsymbol{Y\!}_1)\in A_\epsilon ^{\alpha_1n},\text{and}(v^{\alpha_2n}(1),x_{20}^{\alpha_2 n}(1,1),\boldsymbol{Y\!}_2)\in A_\epsilon ^{\alpha_2n}, \text{and}\;(u^{\alpha_3n}(w_{12}),v^{\alpha_3n}(1),\nonumber\\
&x_{13}^{\alpha_3 n}(w_{13},w_{12},1),x_{23}^{\alpha_3 n}(w_{23},w_{12},1),\boldsymbol{Y\!}_3)\in A_\epsilon ^{\alpha_3n} \;\text{for some}\;w_{12}\neq1 \;\text{and any}\;(w_{10},w_{13},w_{23})\};\nonumber\\
E_{11}:=&\{(u^{\alpha_1n}(1),x_{10}^{\alpha_1 n}(1,1),\boldsymbol{Y\!}_1)\in A_\epsilon ^{\alpha_1n},\text{and}(v^{\alpha_2n}(w_{21}),x_{20}^{\alpha_2 n}(w_{20},w_{21}),\boldsymbol{Y\!}_2)\in A_\epsilon ^{\alpha_2n}, \text{and}
(u^{\alpha_3 n}(1),v^{\alpha_3n}(w_{21}),\nonumber\\
&x_{13}^{\alpha_3 n}(w_{13},1,w_{21}),x_{23}^{\alpha_3 n}(w_{23},1,w_{21}),\boldsymbol{Y\!}_3)\in A_\epsilon ^{\alpha_3 n}\;\text{for some}\;w_{21}\neq1 \;\text{and any}\;(w_{20},w_{13},w_{23})\};\nonumber
\end{align}
\begin{align}\label{Eq:erre}
E_{12}:=&\{(u^{\alpha_1n}(w_{12}),x_{10}^{\alpha_1 n}(w_{10},w_{12}),\boldsymbol{Y\!}_1)\in A_\epsilon ^{\alpha_1n},\;\text{and}\;(v^{\alpha_2n}(w_{21}),x_{20}^{\alpha_2 n}(w_{20},w_{21}),\boldsymbol{Y\!}_2)\in A_\epsilon ^{\alpha_2n}, \:\text{and}\nonumber\\
&(u^{\alpha_3 n}(w_{12}),v^{\alpha_3 n}(w_{21}),x_{13}^{\alpha_3 n}(w_{13},w_{12},w_{21}),x_{23}^{\alpha_3 n}(w_{23},w_{12},w_{21}),\boldsymbol{Y\!}_3)\in A_\epsilon ^{\alpha_3n}\nonumber\\
&\text{for some}\;(w_{12}\neq1,w_{21}\neq1) \;\text{and any}\;(w_{10},w_{13},w_{20},w_{23})\}.
\end{align}
By the Asymptotic Equipartition Property (AEP) \cite{hansen1977mfr}, we have $(P_{E_1},P_{E_3},P_{E_5},P_{E_9})\rightarrow 0$ as $n\rightarrow \infty$. Then, by using the packing lemma \cite{hsce612}, we can easily find that
\begin{itemize}
\item
$P_{E_2}\rightarrow 0$ as $n\rightarrow \infty$ if $R_{10}\leq \alpha_1 I(X_{10};Y_1|U)$.
\item
$P_{E_4}\rightarrow 0$ as $n\rightarrow \infty$ if $R_{20}\leq \alpha_2 I(X_{20};Y_2|V)$.
\item
$P_{E_6}\rightarrow 0$ as $n\rightarrow \infty$ if $R_{13}\leq \alpha_3 I(X_{13};Y_3|U,V,X_{23})$.
\item
$P_{E_7}\rightarrow 0$ as $n\rightarrow \infty$ if $R_{23}\leq \alpha_3 I(X_{23};Y_3|U,V,X_{13})$.
\item
$P_{E_8}\rightarrow 0$ as $n\rightarrow \infty$ if $R_{13}+R_{23}\leq \alpha_3 I(X_{13},X_{23};Y_3|U,V)$.
\end{itemize}
The analysis of the error events $(E_{10},E_{11},E_{12})$ is more complicated because these error events are defined over multiple time slots. Starting with $E_{10}$, we can express the probability of this error event as
\begin{align*}
P_{E_{10}}=\sum_{i=1}^{2^{n(R_{1}+R_{23})}}P_{E_{10i}}\; \text{where}\; P_{E_{10i}}=P_{E_{10i}^1}\times P_{E_{10i}^2}\times P_{E_{10i}^3}
\end{align*}
\noindent where $P_{E_{10i}}$ is the probability of error for a particular set of messages $(w_{10},w_{12},w_{13},1,1,w_{23})$ and $(E_{10i}^1, E_{10i}^2, E_{10i}^3)$ are the events correspond to the first, second, and third time slot, respectively. While $P_{E_{10i}^2}\rightarrow 1$ as $n\rightarrow \infty$ by the AEP, $P_{E_{10i}^1}$ can be bounded as
\noindent
\begin{align*}
P_{E_{10i}^1}=&\sum_{\left(u,x_{10},y_{1}\right)\in A_\epsilon^{\alpha_1n}}p(u)p(x_{10}|u)p(y_1)\\
\leq &2^{\alpha_1n(H(U,X_{10},Y_1)+\epsilon)}2^{-\alpha_1n(H(U,X_{10})-\epsilon)}\cdot2^{-\alpha_1n(H(Y_1)-\epsilon)}\\
=&2^{-\alpha_1n(I(U,X_{10};Y_1)-3\epsilon)}\\
=&2^{-\alpha_1n(I(X_{10};Y_1)-3\epsilon)}.
\end{align*}
\noindent Similarly, $P_{E_{10i}^3}$ can be bounded as
\begin{align*}
P_{E_{10i}^3}=&\sum_{\left(u,v,x_{13},x_{23},y_{3}\right)\in A_\epsilon^{\alpha_3n}}p(u)p(v)p(x_{13}|u,v)p(x_{23}|u,v)p(y_3|v)\\
\leq&\;2^{\alpha_3n(H(U,V,X_{13},X_{23},Y_3)+\epsilon)}2^{-\alpha_3n(H(U,V,X_{13},X_{23})-\epsilon)}2^{-\alpha_3n(H(Y_3|V)-\epsilon)}\\
=&\;2^{-\alpha_3n(I(U,X_{13},X_{23};Y_3|V)-3\epsilon)}\\
=&\;2^{-\alpha_3n(I(X_{13},X_{23};Y_3|V)-3\epsilon)}.
\end{align*}
\noindent Therefore, $P_{E_{10}}$ has the upper bound
\begin{align*}
P_{E_{10}}\leq&2^{n(R_{1}+R_{23})}\cdot2^{-\alpha_1nI(X_{10};Y_1)-\alpha_3nI(X_{13},X_{23};Y_3)}.
\end{align*}
Hence, $P_{E_{10}}\rightarrow0$ as $n\rightarrow\infty$ if $R_{1}+R_{23}\leq\;\alpha_1 I(X_{10};Y_1)+\alpha_3 I(X_{13},X_{23};Y_3|V)$.

Following similar steps with $E_{11}$, we have $P_{E_{1}}\rightarrow0$ as $n\rightarrow\infty$ if $R_{2}+R_{13}\leq\;\alpha_2 I(X_{20};Y_2)+\alpha_3 I(X_{13},X_{23};Y_3|U)$.

Finally, for $E_{12}$, we can express its probability as
\begin{align*}
P_{E_{12}}=\sum_{i=1}^{2^{n\left(R_{1}+R_{2}\right)}-1}P_{E_{12i}},\; \text{where}\; P_{E_{12i}}=P_{E_{12i}^1}\times P_{E_{12i}^2}\times P_{E_{12i}^3}
\end{align*}
Then, $P_{E_{12i}^1}$ and $P_{E_{12i}^2}$ can be derived in a similar way to $P_{E_{10i}^1}$, whereas $P_{E_{12i}^3}$ can be derived as
\begin{align*}
P_{E_{12i}^3}=&\sum_{\left(u,v,x_{13},x_{23},y_{3}\right)\in A_\epsilon^{\alpha_3n}}p(u)p(v)p(x_{13}|u,v)p(x_{23}|u,v)p(y_3)\\
\leq&2^{\alpha_3n(H(U,V,X_{13},X_{23},Y_3)+\epsilon)}2^{-\alpha_3n(H(U,V,X_{13},X_{23})-\epsilon)}2^{-\alpha_3n(H(Y_3)-\epsilon)}\\
=&2^{-\alpha_3n(I(U,V,X_{13},X_{23};Y_3)-3\epsilon)}\\
=&2^{-\alpha_3n(I(X_{13},X_{23};Y_3)-3\epsilon)}.
\end{align*}
\noindent Therefore, $P_{E_{12}}$ can be upper-bounded as
\noindent
\begin{align*}
P_{E_{12}}\leq &2^{n(R_{1}+R_{2})}\cdot2^{-\alpha_1n(I(X_{10};Y_1)-3\epsilon)-\alpha_2n(I(X_{20};Y_2)-3\epsilon)}\cdot2^{-\alpha_3n(I(X_{13},X_{23};Y_3)-3\epsilon)},
\end{align*}

\noindent and thus $P_{E_{12}}\rightarrow0$ as $n\rightarrow\infty$ if $R_1+R_2\leq \alpha_1 I(X_{10};Y_1)+ \alpha_2 I(X_{20};Y_2) + \alpha_3 I(X_{13},X_{23};Y)$. Combining all rate constraints give the rate region in (\ref{Eq:d2dec}).

\section{Error Analysis of the PDF scheme with Separate Decoding}
As in Appendix A, because of the symmetry of the random code generation, without loss of generality, we assume that the message vector $(w_{12}=w_{21}=w_{10}=w_{20}=w_{13}=w_{23}=1)$ was sent. Then, the error events at the destination in the third time slot are given as
\begin{align}\label{Eq:serv3}
E_1:=&\{(u^{\alpha_3 n}(1),v^{\alpha_3 n}(1),x_{13}^{\alpha_3n}(1,1,1),x_{23}^{\alpha_3 n}(1,1,1), \boldsymbol{Y\!}_3)\not\in A_\epsilon ^{\alpha_3 n}\};\nonumber\\
E_2:=&\{(u^{\alpha_3 n}(1),v^{\alpha_3 n}(1),x_{13}^{\alpha_3 n}(w_{13},1,1),x_{23}^{\alpha_3 n}(1,1,1), \boldsymbol{Y\!}_3)\in A_\epsilon ^{\alpha_3 n}\;\text{for some}\;w_{13}\neq1\};\nonumber\\
E_3:=&\{(u^{\alpha_3 n}(1),v^{\alpha_3 n}(1),x_{13}^{\alpha_3 n}(1,1,1),x_{23}^{\alpha_3 n}(w_{23},1,1), \boldsymbol{Y\!}_3)\in A_\epsilon ^{\alpha_3 n}\;\text{for some}\;w_{23}\neq1\};\nonumber\\
E_4:=&\{(u^{\alpha_3 n}(1),v^{\alpha_3 n}(1),x_{13}^{\alpha_3 n}(w_{13},1,1),x_{23}^{\alpha_3 n}(w_{23},1,1), \boldsymbol{Y\!}_3)\in A_\epsilon ^{\alpha_3 n}\;\text{for some}(w_{13}\neq1, w_{23}\neq1)\};\nonumber\\
E_5:=&\{(u^{\alpha_3 n}(1),v^{\alpha_3 n}(1),x_{13}^{\alpha_3 n}(1,1,1),x_{23}^{\alpha_3 n}(1,1,1),\boldsymbol{Y\!}_3)\not\in A_\epsilon ^{\alpha_3 n}\};\nonumber\\
E_{6}:=&\{(u^{\alpha_3n}(w_{12}),v^{\alpha_3n}(1),x_{13}^{\alpha_3 n}(w_{13},w_{12},1),x_{23}^{\alpha_3 n}(w_{23},w_{12},1),\boldsymbol{Y\!}_3)\in A_\epsilon ^{\alpha_3n} \;\text{for some}\;w_{12}\neq1 \;\text{and any}\;(w_{13},w_{23})\};\nonumber\\
E_{7}:=&\{(u^{\alpha_3 n}(1),v^{\alpha_3n}(w_{21}),
x_{13}^{\alpha_3 n}(w_{13},1,w_{21}),x_{23}^{\alpha_3 n}(w_{23},1,w_{21}),\boldsymbol{Y\!}_3)\in A_\epsilon ^{\alpha_3 n}\;\text{for some}\;w_{21}\neq1 \;\text{and any}\;(w_{13},w_{23})\};\nonumber\\
E_{8}:=&\{(u^{\alpha_3 n}(w_{12}),v^{\alpha_3 n}(w_{21}),x_{13}^{\alpha_3 n}(w_{13},w_{12},w_{21}),x_{23}^{\alpha_3 n}(w_{23},w_{12},w_{21}),\boldsymbol{Y\!}_3)\in A_\epsilon ^{\alpha_3n}\nonumber\\
&\text{for some}\;(w_{12}\neq1,w_{21}\neq1) \;\text{and any}\;(w_{13},w_{23})\}.
\end{align}
Similar to Appendix A, we have $(P_{E_1},P_{E_5})\rightarrow 0$ as $n\rightarrow \infty$ by the AEP. By the packing lemma, we have
\begin{itemize}
\item
$P_{E_2}\rightarrow 0$ as $n\rightarrow \infty$ if $R_{13}\leq \alpha_3 I(X_{13};Y_3|U,V,X_{23})$.
\item
$P_{E_3}\rightarrow 0$ as $n\rightarrow \infty$ if $R_{23}\leq \alpha_3 I(X_{23};Y_3|U,V,X_{13})$.
\item
$P_{E_4}\rightarrow 0$ as $n\rightarrow \infty$ if $R_{13}+R_{23}\leq \alpha_3 I(X_{13},X_{23};Y_3|U,V)$.
\item
$P_{E_6}\rightarrow 0$ as $n\rightarrow \infty$ if $R_{12}+R_{13}+R_{23}\leq \alpha_3 I(X_{13},X_{23};Y_3|V)$.
\item
$P_{E_7}\rightarrow 0$ as $n\rightarrow \infty$ if $R_{21}+R_{13}+R_{23}\leq \alpha_3 I(X_{13},X_{23};Y_3|U)$.
\item
$P_{E_8}\rightarrow 0$ as $n\rightarrow \infty$ if $R_{12}+R_{21}+R_{13}+R_{23}\leq \alpha_3 I(X_{13},X_{23};Y_3)$.
\end{itemize}
Now, after decoding the messages $(w_{12},w_{21},w_{13},w_{23})$, the destination goes back to the first two time slots to decode $(w_{10},w_{20})$ assuming that it already decodes $(w_{12},w_{21},w_{13},w_{23})$ correctly. The error events in the first two time slots are

\noindent
\begin{align}\label{Eq:sev12}
E_1:=&\{(u^{\alpha_1 n}(1), x_{10}^{\alpha_1 n}(1,1),\boldsymbol{Y\!}_1)\not\in A_\epsilon ^{\alpha_1 n}\};\nonumber\\
E_2:=&\{(u^{\alpha_1 n}(1), x_{10}^{\alpha_1 n}(w_{10},1),\boldsymbol{Y\!}_1)\in A_\epsilon ^{\alpha_1 n} \;\text{for some}\;w_{10}\neq1\};\nonumber\\
E_3:=&\{(v^{\alpha_2 n}(1), x_{20}^{\alpha_2 n}(1,1),\boldsymbol{Y\!}_2)\not\in A_\epsilon ^{\alpha_2 n}\};\nonumber\\
E_4:=&\{(v^{\alpha_2 n}(1), x_{20}^{\alpha_2 n}(w_{20},1),\boldsymbol{Y\!}_2)\in A_\epsilon ^{\alpha_2 n} \;\text{for some}\;w_{20}\neq1\}.
\end{align}
\noindent Here, $(P_{E_1},P_{E_3})\rightarrow 0$ as $n\rightarrow \infty$ by the AEP. By the packing lemma, $(P_{E_2},P_{E_4})\rightarrow 0$ as $n\rightarrow \infty$ if $R_{10}\leq \alpha_1 I(X_{10};Y_1|U)$ and $R_{20}\leq \alpha_1 I(X_{20};Y_2|V)$, respectively. Hence, we obtain the rate region (\ref{co2rr})

\section{Proof of the Outer Bounds}
In this Appendix, we prove the outer bounds given in Section \ref{sec:out. cap}.
\subsection{Proof of Theorem 3}
We derive an outer bound for the half-duplex MAC-GF with rate constraints similar to the achievable region of the PDF scheme.

Starting with $R_1$, given any sequence of $(2^{nR_1},2^{nR_2},n)$ codes with $P_e\rightarrow 0$, we have
\begin{align}\label{ste}
nR_1=&\;H(W_1)=H(W_1|W_2)\nonumber\\
=&\;I(W_1;Y^n,Y_{12}^n,Y_{21}^n|W_2)+H(W_1|Y^n,Y_{12}^n,Y_{21}^n,W_2)\nonumber\\
\leq&\;I(W_1;Y^n,Y_{12}^n,Y_{21}^n|W_2)+n\epsilon,\;
\end{align}
\noindent where (\ref{ste}) follows from Fano's inequality. Now, let's consider the first part of (\ref{ste}). We have
\begin{align}\label{nde}
I(W_1;Y^n,Y_{12}^n,Y_{21}^n|W_2)=&\;\sum_{i=1}^n I(W_1;Y_i,Y_{12i},Y_{21i}|W_2,Y^{i-1},Y_{12}^{i-1},Y_{21}^{i-1})\nonumber\\
=&\;\sum_{i=1}^{\alpha_1 n} I(W_1;Y_i,Y_{12i}|W_2,Y_{12}^{i-1},Y^{i-1})\nonumber\\
&+\sum_{i=\alpha_1n+1}^{(\alpha_1+\alpha_2) n}I(W_1;Y_i,Y_{21i}|W_2,Y^{\alpha_1n},Y_{\alpha_1n+1}^{i-1},Y_{12}^{\alpha_1n},Y_{21}^{i-1})\nonumber\\
&+\sum_{i=(\alpha_1+\alpha_2)n+1}^{n}\!\!\!\!\!\!\!\!\!I(W_1;Y_i|W_2,Y_{(\alpha_1+\alpha_2)n+1}^{i-1},Y_{12}^{\alpha_1n},Y_{21}^{\alpha_2n})
\end{align}
where $Y_{21}^{\alpha_2n}$ is the second channel output sequence.  Now, the first term in (\ref{nde}) can be bounded as
\begin{align}\label{in1p1}
\sum_{i=1}^{\alpha_1 n} I(W_1;Y_i,Y_{12i}|W_2,Y_{12}^{i-1},Y^{i-1})\stackrel{(a)}{\leq}&\;\sum_{i=1}^{\alpha_1n} I(X_{10i};Y_i,Y_{12i}|W_2,Y^{i-1},Y_{12}^{i-1})\nonumber\\
=&\sum_{i=1}^{\alpha_1n} H(Y_i,Y_{12i}|W_2,Y^{i-1},Y_{12}^{i-1})-H(Y_i,Y_{12i}|X_{10i},W_2,Y^{i-1},Y_{12}^{i-1})\nonumber\\
\stackrel{(b)}{\leq}& \sum_{i=1}^{\alpha_1n} H(Y_i,Y_{12i})-H(Y_i,Y_{12i}|X_{10i})\nonumber\\
=&\;\sum_{i=1}^{\alpha_1n} I(X_{10i};Y_i,Y_{12i})
\end{align}
\noindent where (a) follows from the data processing inequality since $W_1\rightarrow X_{10}\rightarrow (Y_1,Y_{12})$ forms a Markov chain and $(b)$ follows from removing conditioning and the memoryless property of the channel during the first time slot $p(y,y_{12}|x_{10})$.

Moving to the second part of (\ref{nde})
\begin{align}\label{in1p2}
&\sum_{i=\alpha_1n+1}^{(\alpha_1+\alpha_2)n}I(W_1;Y_i,Y_{21i}|W_2,Y^{\alpha_1n},Y^{i-1},Y_{12}^{\alpha_1n},Y_{21}^{i-1})\nonumber\\
\stackrel {(a)}{=}&\sum_{i=\alpha_1n+1}^{(\alpha_1+\alpha_2)n}I(W_1;Y_i,Y_{21i}|X_{20i},W_2,Y^{\alpha_1n},Y^{i-1},Y_{12}^{\alpha_1n},Y_{21}^{i-1})\nonumber\\
=&\sum_{i=\alpha_1n+1}^{(\alpha_1+\alpha_2)n}H(Y_i,Y_{21i}|X_{20i},W_2,Y^{\alpha_1n},Y^{i-1},Y_{12}^{\alpha_1n},Y_{21}^{i-1})-H(Y_i,Y_{21i}|W_1,X_{20i},W_2,Y^{\alpha_1n},Y^{i-1},Y_{12}^{\alpha_1n},Y_{21}^{i-1})\nonumber\\
\stackrel {(b)}{=}&\sum_{i=\alpha_1n+1}^{(\alpha_1+\alpha_2)n}H(Y_i,Y_{21i}|X_{20i})-H(Y_i,Y_{21i}|X_{2i})=0
\end{align}
where (a) follows since in the second time slot, for a given code, $X_{20i}=f_i(W_2)$ and (b) follows from the memoryless property of this channel $p(y,y_{21}|x_{20})$.

Finally, the third part of (\ref{nde}) can be expressed as
\begin{align}\label{rde}
&\sum_{i=(\alpha_1+\alpha_2)n+1}^{n}I(W_1;Y_i|W_2,Y^{i-1},Y_{12}^{\alpha_1n},Y_{21}^{\alpha_2n})\nonumber\\
=&\;\sum_{i=(\alpha_1+\alpha_2)n+1}^{n} H(Y_i|W_2,Y_{12}^{\alpha_1n},Y_{21}^{\alpha_2n},Y^{i-1})-H(Y_i|W_2,Y_{12}^{\alpha_1n},W_1,Y_{21}^{\alpha_2n},Y^{i-1})\nonumber\\
\stackrel {(a)}{=}&\;\sum_{i=(\alpha_1+\alpha_2)n+1}^{n} H(Y_i|X_{23i},W_2,Y_{12}^{\alpha_1n},Y_{21}^{\alpha_2n},Y^{i-1})-H(Y_i|X_{23i},W_2,Y_{12}^{\alpha_1n},X_{13i},W_1,Y_{21i}^{\alpha_2n},Y^{i-1})\nonumber\\
\stackrel {(b)}{\leq}&\;\sum_{i=(\alpha_1+\alpha_2)n+1}^{n} H(Y_i|X_{23i},Y_{12}^{\alpha_1n},Y_{21}^{\alpha_2n})-H(Y_i|X_{23i},Y_{12}^{\alpha_1n},X_{13i},Y_{21i}^{\alpha_2n})\nonumber\\
=&\;\sum_{i=(\alpha_1+\alpha_2)n+1}^{n}  I(X_{13i};Y_i|X_{23i},Y_{12}^{\alpha_1n},Y_{21}^{\alpha_2})\nonumber\\
=&\;\sum_{i=(\alpha_1+\alpha_2)n+1}^{n}  I(X_{13i};Y_i|X_{23i},U,V)
\end{align}
where $U=Y_{12}^{\alpha_1n}$ and $V=Y_{21}^{\alpha_2 n}$; (a) follows from $X_{13i}=f_{1i}(W_1,Y_{21}^{\alpha_2n})$, $X_{23i}=f_{2i}(W_2,Y_{12}^{\alpha_1 n})$ and $(W_1,W_2,Y^{i-1},Y_{21}^{i-1},Y_{12}^{i-1})\rightarrow (X_{1i}, X_{2i})\rightarrow Y_i$ forms a Markov chain;
(b) follows from removing conditioning and the memoryless property of the channel $p(y|x_{13},x_{23})$.

Thus, from (\ref{in1p1}), (\ref{in1p2}) and (\ref{rde}), we have
\begin{align}\label{bind1}
nR_1\leq \sum_{i=1}^{\alpha_1n} I(X_{10i};Y_{1i},Y_{12i})+\sum_{i=(\alpha_1+\alpha_2)n+1}^{n} I(X_{13i};Y_{3i}|X_{23i},U,V)+n\epsilon.
\end{align}
Similarly,
\begin{align}\label{bind2}
nR_2\leq \sum_{i=\alpha_1n+1}^{(\alpha_1+\alpha_2)n} I(X_{20i};Y_{2i},Y_{21i})+\sum_{i=(\alpha_1+\alpha_2)n+1}^{n} I(X_{23i};Y_{3i}|X_{1i},U,V)+n\epsilon.
\end{align}

Moving to the sum rate, based on Fano's inequality, we have
\begin{align}\label{fthe}
n(R_1+R_2)=&\;H(W_1,W_2)\nonumber\\
=&\;I(W_1,W_2;Y^n,Y_{12}^n,Y_{21}^n)+H(W_1,W_2|Y^n,Y_{12}^n,Y_{21}^n)\nonumber\\
\leq&\;I(W_1,W_2;Y^n,Y_{12}^n,Y_{21}^n)+n\epsilon.\;
\end{align}
The first term in (\ref{fthe}) can be bounded as
\begin{align}\label{su1p}
I(W_1,W_2;Y^n,Y_{12}^n,Y_{21}^n)=&\sum_{i=1}^n I(W_1,W_2;Y_i,Y_{12i},Y_{21i}|Y^{i-1},Y_{12}^{i-1},Y_{21}^{i-1})\nonumber\\
=&\;\sum_{i=1}^{\alpha_1 n} I(W_1,W_2;Y_i,Y_{12i}|Y_{12}^{i-1},Y^{i-1})\nonumber\\
&+\sum_{i=\alpha_1n+1}^{(\alpha_1+\alpha_2)n} I(W_1,W_2;Y_i,Y_{21i}|Y_{12}^{\alpha_1n},Y_{21}^{i-1},Y^{i-1})\nonumber\\
&+\sum_{i=(\alpha_1+\alpha_2)n+1}^{n}\!\!\!\!\!\!\!\!I(W_1,W_2;Y_i|Y^{i-1},Y_{12}^{\alpha_1n},Y_{21}^{\alpha_2n})
\end{align}
The first part of (\ref{su1p}) can be bounded as
\begin{align}
&\sum_{i=1}^{\alpha_1 n} I(W_1,W_2;Y_i,Y_{12i}|Y_{12}^{i-1},Y^{i-1})\nonumber\\
&=\sum_{i=1}^{\alpha_1 n} H(Y_i,Y_{12i}|Y_{12}^{i-1},Y^{i-1})-H(Y_i,Y_{12i}|W_1,W_2,Y_{12}^{i-1},Y^{i-1})\nonumber\\
&\stackrel{(a)}{=}\sum_{i=1}^{\alpha_1 n} H(Y_i,Y_{12i}|Y_{12}^{i-1},Y^{i-1})-H(Y_i,Y_{12i}|X_{10i},W_1,W_2,Y_{12}^{i-1},Y^{i-1})\nonumber\\
&\stackrel{(b)}{\leq}\sum_{i=1}^{\alpha_1 n} H(Y_i,Y_{12i})-H(Y_i,Y_{12i}|X_{10i})\nonumber\\
&= \sum_{i=1}^{\alpha_1 n} I(X_{10i};Y_i,Y_{12i})
\end{align}
where (a) follows since for the channel in the first time slot, for a given code, $X_{10i}=f_i(W_1)$; (b) follows from the memoryless property of the channel $p(y,y_{12}|x_{10})$ and from removing conditioning.

Similarly, the second part of (\ref{su1p}) can be bounded as
\begin{align}\label {su1p2}
&\sum_{i=\alpha_1n+1}^{(\alpha_1+\alpha_2)n} I(W_1,W_2;Y_i,Y_{21i}|Y_{12}^{\alpha_1n},Y_{21}^{i-1},Y^{i-1})\nonumber\\
&=\sum_{i=\alpha_1n+1}^{(\alpha_1+\alpha_2)n}H(Y_i,Y_{21i}|Y_{12}^{\alpha_1n},Y_{21}^{i-1},Y^{i-1})-H(Y_i,Y_{21i}|W_1,W_2,Y_{12}^{\alpha_1n},Y_{21}^{i-1},Y^{i-1})\nonumber\\
&\stackrel{(a)}{=}\sum_{i=\alpha_1n+1}^{(\alpha_1+\alpha_2)n}H(Y_i,Y_{21i}|Y_{12}^{\alpha_1n},Y_{21}^{i-1},Y^{i-1})-H(Y_i,Y_{21i}|W_1,X_{20i},W_2,Y_{12}^{\alpha_1n},Y_{21}^{i-1},Y^{i-1})\nonumber\\
&\stackrel{(b)}{\leq}\sum_{i=\alpha_1n+1}^{(\alpha_1+\alpha_2)n}H(Y_i,Y_{21i})-H(Y_i,Y_{21i}|X_{20i})
=\sum_{i=\alpha_1n+1}^{(\alpha_1+\alpha_2)n}I(X_{20i},Y_i,Y_{21i})
\end{align}
where (a) follows since for the channel in the second time slot, for a given code, $X_{20i}=f_i(W_2)$; (b) follows from the memoryless property of the channel $p(y,y_{21}|x_{20})$ and from removing conditioning.

Moving to the third part of (\ref{su1p}):
\begin{align}\label {su1p3}
&\sum_{i=(\alpha_1+\alpha_2)n+1}^{n}\!\!\!\!\!\!\!\!I(W_1,W_2;Y_i|Y^{i-1},Y_{12}^{\alpha_1n},Y_{21}^{\alpha_2n})\nonumber\\
&=\sum_{i=(\alpha_1+\alpha_2)n+1}^{n}\!\!\!\!\!\!\!\!H(Y_i|Y^{i-1},Y_{12}^{\alpha_1n},Y_{21}^{\alpha_2n})-H(Y_i|Y^{i-1},W_2,Y_{12}^{\alpha_1n},W_1,Y_{21}^{\alpha_2n})\nonumber\\
&\stackrel{(a)}{=}\sum_{i=(\alpha_1+\alpha_2)n+1}^{n}\!\!\!\!\!\!\!\!H(Y_i|Y^{i-1},Y_{12}^{\alpha_1n},Y_{21}^{\alpha_2n})-H(Y_i|Y^{i-1},X_{23i},W_2,Y_{12}^{\alpha_1n},X_{13i},W_1,Y_{21}^{\alpha_2n})\nonumber\\
&\stackrel{(b)}{=}\sum_{i=(\alpha_1+\alpha_2)n+1}^{n}\!\!\!\!\!\!\!\!H(Y_i|Y^{i-1},Y_{12}^{\alpha_1n},Y_{21}^{\alpha_2n})-H(Y_i|X_{23i},Y_{12}^{\alpha_1n},X_{13i},Y_{21}^{\alpha_2n})\nonumber\\
&\stackrel{(c)}{\leq} \sum_{i=(\alpha_1+\alpha_2)n+1}^{n}\!\!\!\!\!\!\!\!H(Y_i|Y_{12}^{\alpha_1n},Y_{21}^{\alpha_2n})-H(Y_i|X_{23i},Y_{12}^{\alpha_1n},X_{13i},Y_{21}^{\alpha_2n})\nonumber\\
&=\sum_{i=(\alpha_1+\alpha_2)n+1}^{n}I(X_{13i},X_{23i};Y_i|Y_{12}^{\alpha_1n},Y_{21}^{\alpha_2n})\nonumber\\
&=\sum_{i=(\alpha_1+\alpha_2)n+1}^{n}I(X_{13i},X_{23i};Y_i|U,V)
\end{align}
Here (a) follows since for the channel in the third time slot, for a given code, $X_{13i}=f_{1i}(W_1,Y_{21}^{\alpha_2n})$, $X_{23i}=f_{2i}(W_2,Y_{12}^{\alpha_1 n})$; (b) follows from removing $Y^{i-1},W_1,W_2$ since in this channel,  $(W_1,W_2,Y^{i-1},Y_{21}^{i-1},Y_{12}^{i-1})\rightarrow (X_{1i}, X_{2i})\rightarrow Y_i$ forms a Markov chain; (c) follows from removing conditioning.

Thus, from (\ref {su1p}), (\ref{su1p2}) and (\ref{su1p3}), we have
\begin{align}\label{bsu1}
n(R_1+R_2)\leq\;\sum_{i=1}^{\alpha_1 n} I(X_{10_i};Y_{1i},Y_{12i})+\!\!\sum_{i=\alpha_1n+1}^{(\alpha_1+\alpha_2)n} I(X_{20i};Y_{2i},Y_{21i})+\!\!\sum_{i=(\alpha_1+\alpha_2)n+1}^{n}\!\!\!\!\!\!\!\!I(X_{13i},X_{23i};Y_i|U,V)+n\epsilon.
\end{align}

Another bound for the sum rate can be derived as
\begin{align}\label{su2}
n(R_1+R_2)=&\;H(W_1,W_2)\nonumber\\
=&\;I(W_1,W_2;Y^n,Y_{21}^n)+H(W_1,W_2|Y^n,Y_{21}^n)\nonumber\\
\leq&\;I(W_1,W_2;Y^n,Y_{21}^n)+n\epsilon\;
\end{align}
where (\ref{su2}) follows from the Fano's inequality. The first term in (\ref{su2}) can be bounded as
\begin{align}\label {su2pp}
&I(W_1,W_2;Y^n,Y_{21}^n)\nonumber\\
&=\;\sum_{i=1}^{\alpha_1 n} I(W_1,W_2;Y_i|Y^{i-1})+\sum_{i=\alpha_1n+1}^{(\alpha_1+\alpha_2)n} I(W_1,W_2;Y_i,Y_{21i}|Y_{21}^{i-1},Y^{i-1})+\sum_{i=(\alpha_1+\alpha_2)n+1}^{n}\!\!\!\!\!\!\!\!I(W_1,W_2;Y_i|Y^{i-1},Y_{21}^{\alpha_2n})
\end{align}
Following similar lines of argument, the first part of (\ref{su2pp}) can be bounded as
\begin{align}
\sum_{i=1}^{\alpha_1 n} I(W_1,W_2;Y_i|Y^{i-1})=&\;\sum_{i=1}^{\alpha_1 n} I(W_1;Y_i|Y^{i-1})+I(W_2;Y_i|Y^{i-1},W_1)\nonumber
\end{align}
\begin{align}\label{su2p1}
\stackrel{(a)}{=}&\;\sum_{i=1}^{\alpha_1 n} I(W_1;Y_i|Y^{i-1})+I(W_2;Y_i|Y^{i-1},W_1,X_{10i})\nonumber\\
\stackrel{(b)}{\leq}&\;\sum_{i=1}^{\alpha_1 n} I(X_{10i};Y_i|Y^{i-1})\nonumber\\
\stackrel{(c)}{\leq}&\;\sum_{i=1}^{\alpha_1 n} I(X_{10i};Y_i)
\end{align}
where (a) follows since for the channel in the first time slot, for a given code, $X_{10i}=f_i(W_1)$, (b) follows from the Markov chain $(W_2,Y^{i-1},W_1)\rightarrow X_{10i}\rightarrow (Y_i,Y_{12i})$, and (c) follows from removing conditioning and the same Markov chain.

Similarly, the second part of (\ref{su2pp}) can be bounded as
\begin{align}\label {su2p2}
\sum_{i=\alpha_1n+1}^{(\alpha_1+\alpha_2)n}\!\!\!\!I(W_1,W_2;Y_i,Y_{21i}|Y_{21}^{i-1},Y^{i-1})=&\;\sum_{i=\alpha_1n+1}^{(\alpha_1+\alpha_2)n}\!\!\!\!I(W_2;Y_i,Y_{21i}|Y_{21}^{i-1},Y^{i-1})+I(W_1;Y_i,Y_{21i}|W_2,Y_{21}^{i-1},Y^{i-1})\nonumber\\
\stackrel{(a)}{=}&\;\sum_{i=\alpha_1n+1}^{(\alpha_1+\alpha_2)n}\!\!\!\!I(W_2;Y_i,Y_{21i}|Y_{21}^{i-1},Y^{i-1})\nonumber\\
\stackrel{(b)}{\leq}&\;\sum_{i=\alpha_1n+1}^{(\alpha_1+\alpha_2)n}\!\!\!\!I(X_{20i};Y_i,Y_{21i}|Y_{21}^{i-1},Y^{i-1})\nonumber\\
\stackrel{(c)}{\leq}&\;\sum_{i=\alpha_1n+1}^{(\alpha_1+\alpha_2)n}\!\!\!\!I(X_{20i};Y_i,Y_{21i})
\end{align}
where (a) follows since $I(W_2;Y_i,Y_{21i}|Y_{21}^{i-1},Y^{i-1})=0$ as shown in (\ref{in1p2}), (b) follows from the Markov chain $(W_2,Y^{i-1},Y_{21}^{i-1})\rightarrow X_{20i}\rightarrow (Y_i,Y_{21i})$ in the channel of the second time slot, and (c) from removing conditioning and the same Markov chain.

Moving to the last part of (\ref{su2pp}), we have
\begin{align}
&\sum_{i=(\alpha_1+\alpha_2)n+1}^{n}\!\!\!\!\!\!\!\!I(W_1,W_2;Y_i|Y^{i-1},Y_{21}^{\alpha_2n})\nonumber\\
&=\sum_{i=(\alpha_1+\alpha_2)n+1}^{n}\!\!\!\!\!\!\!\!H(Y_i|Y^{i-1},Y_{21}^{\alpha_2n})-H(Y_i|W_1,W_2,Y^{i-1},Y_{21}^{\alpha_2n})\nonumber\\
&\stackrel{(a)}{\leq}\sum_{i=(\alpha_1+\alpha_2)n+1}^{n}\!\!\!\!\!\!\!\!H(Y_i|Y^{i-1},Y_{21}^{\alpha_2n})-H(Y_i|W_1,W_2,Y^{i-1},Y_{21}^{\alpha_2n},Y_{12}^{\alpha_1n})\nonumber\\
&\stackrel{(b)}{=}\sum_{i=(\alpha_1+\alpha_2)n+1}^{n}\!\!\!\!\!\!\!\!H(Y_i|Y^{i-1},Y_{21}^{\alpha_2n})-H(Y_i|X_{13i},X_{23i},W_1,W_2,Y^{i-1},Y_{21}^{\alpha_2n},Y_{12}^{\alpha_1n})\nonumber\\
&\stackrel{(c)}{=}\sum_{i=(\alpha_1+\alpha_2)n+1}^{n}\!\!\!\!\!\!\!\!H(Y_i|Y^{i-1},Y_{21}^{\alpha_2n})-H(Y_i|X_{13i},X_{23i})\nonumber\\
&\stackrel{(d)}{\leq} \sum_{i=(\alpha_1+\alpha_2)n+1}^{n}\!\!\!\!\!\!\!\!H(Y_i|Y_{21}^{\alpha_2n})-H(Y_i|X_{13i},X_{23i},Y_{21}^{\alpha_2n})\nonumber
\end{align}
\begin{align}\label{su2p3}
&=\sum_{i=(\alpha_1+\alpha_2)n+1}^{n}\!\!\!\!\!\!\!\!I(X_{13i},X_{23i};Y_i|Y_{21}^{\alpha_2n})\nonumber\\
&=\sum_{i=(\alpha_1+\alpha_2)n+1}^{n}\!\!\!\!\!\!\!\!I(X_{13i},X_{23i};Y_i|V)
\end{align}
where (a) follows from adding $Y_{12}^{\alpha_1n}$ and that conditioning reduces entropy; (b) and (c) follow since for the channel in the third time slot, for a given code, we have $X_{13i}=f_{1i}(W_1,Y_{21}^{\alpha_2n})$, $X_{23i}=f_{2i}(W_2,Y_{12}^{\alpha_1 n})$ and $(W_1,W_2,Y^{i-1},Y_{21}^{i-1},Y_{12}^{i-1})\rightarrow (X_{1i}, X_{2i})\rightarrow Y_i$ forms a Markov chain; (d) follows from removing conditioning.

Thus, from (\ref{su2p1}), (\ref{su2p2}) and (\ref{su2p3}), we have
\begin{align}\label{bsu2}
n(R_1+R_2)&\leq\;\sum_{i=1}^{\alpha_1 n} I(X_{10_i};Y_{1i})+\sum_{i=\alpha_1n+1}^{(\alpha_1+\alpha_2)n} I(X_{20i};Y_{2i},Y_{21i})+\sum_{i=(\alpha_1+\alpha_2)n+1}^{n}\!\!\!\!\!\!\!\!I(X_{13i},X_{23i};Y_i|V)+n\epsilon.
\end{align}
Similarly, the sum rate can be bounded as
\begin{align}\label{bsu3}
n(R_1+R_2)\leq\;\sum_{i=1}^{\alpha_1 n} I(X_{10_i};Y_{1i},Y_{12i})+\!\!\sum_{i=\alpha_1n+1}^{(\alpha_1+\alpha_2)n} I(X_{20i};Y_{2i})+\!\!\sum_{i=(\alpha_1+\alpha_2)n+1}^{n}\!\!\!\!\!\!\!\!I(X_{13i},X_{23i};Y_i|U)+n\epsilon.
\end{align}

Finally, following the standard converse (for the cut-set bound), we get
\begin{align}\label{bsu4}
n(R_1+R_2)=&\;H(W_1,W_2)\nonumber\\
=&\;I(W_1,W_2;Y^n)+H(W_1,W_2|Y^n)\nonumber\\
\leq&\;I(W_1,W_2;Y^n)+n\epsilon\; \nonumber\\
\leq& I(X_1^n,X_2^n;Y^n)+n\epsilon\nonumber\\
\leq&\;\sum_{i=1}^{\alpha_1 n} I(X_{10_i};Y_{1i})+\sum_{i=\alpha_1n+1}^{(\alpha_1+\alpha_2)n} I(X_{20i};Y_{2i})+\sum_{i=(\alpha_1+\alpha_2)n+1}^{n}\!\!\!\!\!\!\!\!I(X_{13i},X_{23i};Y_i)+n\epsilon.
\end{align}

Now, from the rate constraints (\ref{bind1}), (\ref{bind2}), (\ref{bsu1}), (\ref{bsu2}), (\ref{bsu3}), and (\ref{bsu4}) and after defining a time-sharing random variable $Q$ independent of $(W_1,W_2,X_1^n,X_2^n,U,V,Y^n)$ and uniformly distributed over $[1:n]$, an outer bound for the half-duplex MAC-GF can be written as
\begin{align*}
R_1\leq&\; \alpha_1I(X_{10};Y_1,Y_{12}|Q)+\alpha_3 I(X_{13};Y_3|X_{23},U,V,Q)\\
R_2\leq&\; \alpha_2I(X_{20};Y_2,Y_{21}|Q)+\alpha_3 I(X_{23};Y_3|X_{13},U,V,Q)\\
R_1+R_2\leq&\; \alpha_1I(X_{10};Y_1,Y_{12}|Q)+\alpha_2I(X_{21};Y_2,Y_{21}|Q)+\alpha_3 I(X_{13},X_{23};Y_3|U,V,Q)\\
R_1+R_2\leq&\; \alpha_1I(X_{10};Y_1|Q)+\alpha_2I(X_{20};Y_2,Y_{21}|Q)+\alpha_3 I(X_{13},X_{23};Y_3|V,Q)\\
R_1+R_2\leq&\; \alpha_1I(X_{10};Y_1,Y_{12}|Q)+\alpha_2I(X_{20};Y_2|Q)+\alpha_3 I(X_{13},X_{23};Y_3|U,Q)\\
R_1+R_2\leq&\; \alpha_1I(X_{10};Y_1|Q)+\alpha_2I(X_{20};Y_2|Q)+\alpha_3 I(X_{13},X_{23};Y_3|Q)
\end{align*}
for some joint distribution $p(q)p(x_{10},u|q)p(x_{20},v|q)p(x_{13}|u,v,q)p(x_{23}|u,v,q)P^{\bullet}$. Since $I(X_{10};Y_1,Y_{12}|Q)\leq I(X_{10};Y_1,Y_{12})$ and the same holds for all other mutual information terms, we get the bound given in Theorem 3.
\subsection{Proof of Corollary 2}
As mentioned in Section \ref{sec:out. cap}, with small modifications on the previous outer bound, we get an outer bound similar to the achievable region of the DF scheme.

By setting $X_{10i}=X_{12i}$, $X_{20i}=X_{21i}$, and $S=(Y_{12}^{\alpha_1n},Y_{21}^{\alpha_2n})=(U,V)$, then the individual rate constraints, the first and last sum rate constraints in (\ref{outeq1}) are kept unchanged. However, the two middle sum rate constraints need minor change.  The last part in (\ref{su2pp}) was bounded as in (\ref{su2p3}). We can bounded it further by removing the conditioning and applying the Markov chain $(W_1,W_2,Y^{i-1},Y_{21}^{i-1},Y_{12}^{i-1})\rightarrow (X_{1i}, X_{2i})\rightarrow Y_i$, as follows.

\begin{align}\label {simsu2p3}
&\sum_{i=(\alpha_1+\alpha_2)n+1}^{n}\!\!\!\!\!\!\!\!I(W_1,W_2;Y_i|Y^{i-1},Y_{21}^{\alpha_2n})\nonumber\\
&\leq \sum_{i=(\alpha_1+\alpha_2)n+1}^{n}\!\!\!\!\!\!\!\!H(Y_i|Y_{21}^{\alpha_2n})-H(Y_i|X_{13i},X_{23i},Y_{21}^{\alpha_2n})\nonumber\\
&\leq\sum_{i=(\alpha_1+\alpha_2)n+1}^{n}\!\!\!\!\!\!\!\!H(Y_i)-H(Y_i|X_{13i},X_{23i})\nonumber\\
&=\sum_{i=(\alpha_1+\alpha_2)n+1}^{n}\!\!\!\!\!\!\!\!I(X_{13i},X_{23i};Y_i).
\end{align}
Similar steps can be used for the third sum constraint in (\ref{outeq1}). Although these changes makes the two middle sum rate constraints similar to those in the DF scheme, they are redundant because they become greater than the last sum rate constraint and hence are removed.
\subsection{Relation with the Dependence Balance Outer Bound for the Full-Duplex MAC-GF}
In \cite{dbout}, the following outer bound is proposed for full-duplex MAC-GF:

\textbf{\emph{Outer Bound for full-duplex MAC-GF }}[Tandon and Ulukus]: \emph{An outer bound of the full-duplex MAC-GF consists of the union of all rate pairs $(R_1,R_2)$ satisfying }
\begin{align}\label{outeqf}
R_1\leq&\; I(X_1;Y,Y_{12}|X_2,S)\nonumber\\
R_2\leq&\; I(X_2;Y,Y_{21}|X_1,S)\nonumber\\
R_1+R_2\leq&\; I(X_1,X_2;Y,Y_{12},Y_{21}|S)\nonumber\\
R_1+R_2\leq&\;  I(X_1,X_2;Y)
\end{align}
\emph{for some joint distribution $p(x_1,x_2,s)p(y|x_1,x_2)p(y_{12}|x_1)p(y_{21}|x_2)$. The rates also satisfy the dependence balance bound $I(X_1,X_2|S)\leq\; I(X_1;X_2|Y_{12},Y_{21},S)$.}

In the proof of this dependence balance outer bound in (\cite{dbout}, Theorem 4), the individual rate $nR_1$ was bounded as $\sum_{i=1}^n I(X_{1i};Y_i,Y_{12i}|X_{2i},Y_{12}^{i-1},Y_{21}^{i-1})$. We can expand this bound over three time slots as
\begin{align}\label{oudep}
nR_1\leq&\;\sum_{i=1}^n I(X_{1i};Y_i,Y_{12i}|X_{2i},Y_{12}^{i-1},Y_{21}^{i-1})\nonumber\\
\stackrel{(a)}{=}&\; \sum_{i=1}^{\alpha_1 n} I(X_{10i};Y_{1i},Y_{12i}|Y_{12}^{i-1}) +\sum_{i=(\alpha_1+\alpha_2)n+1}^{n}\!\!\!\!\!\!\!\!I(X_{13i};Y_{3i}|X_{23i},Y_{12}^{\alpha_1 n},Y_{21}^{\alpha_2n})\nonumber\\
\stackrel{(b)}{\leq}&\;\sum_{i=1}^{\alpha_1 n} I(X_{10i};Y_{1i},Y_{12i}) +\sum_{i=(\alpha_1+\alpha_2)n+1}^{n}\!\!\!\!\!\!\!\!I(X_{13i};Y_{3i}|X_{23i},S)
\end{align}
\noindent where (a) follows because for the channel in the first time slot, $X_{1i}=X_{10i},$ $Y_{i}=Y_{1i},$ for the channel in the second time slot $X_{2i}=Y_{21}^{i-1}=\phi$, $X_{1i}=\phi$ and for the channel in the third time slot, $X_{1i}=X_{13i},$ $X_{2i}=X_{23i},$ $Y_{i}=Y_{3i},$ $Y_{12}^{i-1}=Y_{12}^{\alpha_1 n},$ and $Y_{21}^{i-1}=Y_{21}^{\alpha_2 n}$; (b) follows by setting $S=(Y_{12}^{\alpha_1 n},Y_{21}^{\alpha_2n})$, from removing conditioning and from the Markov chain $Y_{12}^{i-1}\rightarrow X_{10i}\rightarrow (Y_{1i},Y_{12i})$.

Following similar steps with the other constraints in (\ref{outeqf}) and after defining the time sharing random variable $Q$ uniformly distributed over $[1:n]$ and independent of all other random variables, we get the same constraints as those given in Corollary $2$.

However, we can show that for the half-duplex MAC-GF, the dependence balance constraint is automatically satisfied. Starting from the formula of dependence balance constraint given in \cite{dbout,ray4}, we have
\begin{align}\label{depder}
0\leq&\;\sum_{i=1}^n \left(I(X_{1i};X_{2i}|Y_{12i},Y_{21i},Y_{12}^{i-1},Y_{21}^{i-1})-I(X_{1i};X_{2i}|Y_{12}^{i-1},Y_{21}^{i-1})\right)\nonumber\\
=&\;\sum_{i=1}^{\alpha_1n} \left(I(X_{12i};X_{21i}|Y_{12i},Y_{21i},Y_{12}^{i-1},Y_{21}^{i-1})-I(X_{12i};X_{21i}|Y_{12}^{i-1},Y_{21}^{i-1})\right)\nonumber\\
+&\;\sum_{i=\alpha_1n+1}^{(\alpha_1+\alpha_2)n} \!\!\!\!\!\left(I(X_{12i};X_{21i}|Y_{12i},Y_{21i},Y_{12}^{\alpha_1n},Y_{21}^{i-1})-I(X_{12i};X_{21i}|Y_{12}^{\alpha_1n},Y_{21}^{i-1})\right)\nonumber\\
+&\;\sum_{i=\alpha_1n+1}^{(\alpha_1+\alpha_2)n} \!\!\!\!\!\left(I(X_{13i};X_{23i}|Y_{12i},Y_{21i},Y_{12}^{\alpha_1n},Y_{21}^{\alpha_2n})-I(X_{13i};X_{23i}|Y_{12}^{\alpha_1n},Y_{21}^{\alpha_2n})\right)\nonumber\\
\stackrel{(a)}{=}&\;0.
\end{align}
\noindent where (a) follows since $X_{21},$ $X_{12},$ and $(Y_{12},Y_{21})$ are equal to $\phi$ for the channel in the
first, second, and third time slot, respectively. Thus, the dependence balance constraint is
automatically satisfied for the half-duplex MAC-GF.
\section{An Optimal Input Distribution the for Gaussian Channel}
For the discrete-time model of the Gaussian channel given in (\ref{Gchm}), we need to find the optimal input distribution that maximizes the rate region given in (\ref{th1rr}). We will maximize each term in (\ref{th1rr}) individually and then show that jointly Gaussian distribution is an optimal distribution.

Starting with $I(X_{10};Y_{12})$, we have
\begin{align}\label{Dmax}
I(X_{10};Y_{12})=&\;h(Y_{12})-h(Y_{12}|X_{10})=h(Y_{12})-h(Z_1).
\end{align}
\noindent By the maximum entropy theorem, (\ref{Dmax}) is maximized when $Y_{12}$ is Gaussian and since $Y_{12}=K_{12}X_{10}+Z_1$, $X_{10}$ must be Gaussian. Because of superposition encoding, we also have
$I(X_{10};Y_{12})=I(U,X_{10};Y_{12})=I(U;Y_{12})+I(X_{10};Y_{12}|U)$. Now,
\begin{align*}
I(U;Y_{12})=&\;h(Y_{12})-h(Y_{12}|U)\\
\leq&\; h(Y_{12}^G)-h(Y_{12}|U)\\
\leq&\; h(Y_{12}^G)-0.5\text{log}\left(2^{2h(X_{10}|U)}+2^{2h(Z_1)}\right)
\end{align*}
\noindent where $Y_{12}^G$ denotes $Y_{12}$ when $X_{10}$ is Gaussian, and the last inequality follows from the entropy power inequality (EPI). The equality holds when $X_{10}|U$ is Gaussian. For $I(X_{10};Y_{12}|U)$, we have
\begin{align}\label{Dmax3}
I(X_{10};Y_{12}|U)=&\;h(Y_{12}|U)-h(Y_{12}|X_{10},U)\nonumber\\
\leq&\; h(K_{12}X_{10}|U+Z_1)-h(Z_1)
\end{align}
\noindent where the equality holds again when $X_{10}|U$ is Gaussian. Therefore, we conclude that $I(X_{10};Y_{12})$ is
maximized when $(X_{10},U)$ are jointly Gaussian. Similarly, $I(X_{10};Y_1)$ is maximized with the same distribution and $(I(X_{20};Y_{21}),\;I(X_{20};Y_{2}))$ are maximized when $(X_{20},V)$ are jointly Gaussian.

Following similar steps to (\ref{Dmax3}), we can show that
\begin{itemize}
\item
$I(X_{13};Y_3|X_{23},U,V)$ is maximized when $(X_{13}|U,V)$ is Gaussian.
\item
$I(X_{23};Y_3|X_{13},U,V)$ is maximized when $(X_{23}|U,V)$ is Gaussian.
\item
$I(X_{13},X_{23};Y_3|U,V)$ is maximized when $(X_{13}|U,V+X_{23}|U,V)$ is Gaussian.
\item
$I(X_{13},X_{23};Y_3|U)$ is maximized when $(X_{13}|U+X_{23}|U)$ is Gaussian.
\item
$I(X_{13},X_{23};Y_3|V)$ is maximized when $(X_{13}|V+X_{23}|V)$ is Gaussian.
\item
$I(X_{13},X_{23};Y_3)$ is maximized when $(X_{13}+X_{23})$ is Gaussian.
\end{itemize}
Therefore, we conclude that jointly Gaussian distribution for $(X_{13},X_{23},U,V)$ maximizes all above
mutual information expressions. Thus, the rate region in Theorem $1$ is maximized with jointly Gaussian distribution $(X_{10},X_{20},X_{13},$ $X_{23},U,V)\sim N(0,\Sigma),$ where $\Sigma$ is the covariance matrix.

For the input distribution given in Theorem $1$, $(p(x_{10},u)p(x_{20}|v)p(x_{13}|u,v)p(x_{23}|u,v))$, the covariance matrix $\Sigma$ can be expressed as
\begin{align}\label{Ginpd}
\Sigma=\text{cov}(X_{10},X_{20},X_{13},X_{23},U,V)=\begin{bmatrix}
 \tilde{P}_{10} & 0 & \rho_1 & \rho_2 & P_U & 0 \\
 0 & \tilde{P}_{20} & \rho_3 & \rho_4 & 0 & P_V \\
 \rho_1 & \rho_3 & \tilde{P}_{13} & \rho_5 & \rho_1 & \rho_3 \\
 \rho_2 & \rho_4 & \rho_5 & \tilde{P}_{23} & \rho_2 & \rho_4 \\
 P_U & 0 & \rho_1 & \rho_2 & P_U & 0\\
 0 & P_V & \rho_3 & \rho_4 & 0 & P_V
\end{bmatrix}
\end{align}
where
\begin{itemize}
\item
$\tilde{P}_{10}=P_{10}+P_U$, and $\tilde{P}_{20}=P_{20}+P_V$
\item
$\tilde{P}_{13}=P_{13}+c_2P_U+c_3P_V$, and $\tilde{P}_{23}=P_{23}+d_3P_U+d_2P_V$
\item
$\rho_1=\sqrt{c_2}P_U$ and $\rho_2=\sqrt{d_3}P_U$
\item
$\rho_3=\sqrt{c_3}P_V$ and $\rho_4=\sqrt{d_2}P_V$
\item
$\rho_5=\sqrt{c_2d_3}P_U+\sqrt{c_3d_2}P_V$.
\end{itemize}

\section{Equivalence Between the PDF Scheme and the DF Scheme for Gaussian channel}
In order to show the equivalence between these two schemes, we need to show that $R_{\text{PDF}}\subseteq R_{\text{DF}}$ and $R_{\text{DF}}\subseteq R_{\text{PDF}}$. We follow a procedure similar to the interference channel in \cite{hkintf}.

First, to show that $R_{\text{PDF}}\subseteq R_{\text{DF}}$, from the rate regions of the DF and the PDF schemes, we can apply a simple one-to-one mapping as
\begin{itemize}
\item
$P_{12}=P_{10}+P_{U}$ and $P_{21}=P_{20}+P_{V}$
\item
$P_{S_1}=c_2P_{U}+c_3P_{V}$ and $P_{S_2}=d_3P_{U}+d_2P_{V}$
\end{itemize}
Then, the two rate regions will be virtually maximized over the same input probability distribution $P_G^{\ast}$ which is the set of all jointly Gaussian distributions  with covariance matrix given in (\ref{Ginpd}). The distribution for the DF scheme can be obtained from (\ref{Ginpd}) by setting $S=(U,V)$. For a given $P_G^{\ast}$, the two rate regions will have the same power and rate constraints, except the two middle sum rates for which the DF scheme is bigger. Hence, from the rate regions expressions, it can be directly inferred that $R_{\text{PDF}}\subseteq R_{\text{DF}}$.

Now, to show that $R_{\text{DF}}\subseteq R_{\text{PDF}}$, we use the following Corollary:

\textbf{Corollary $8$.} \emph{For a given jointly Gaussian input distribution $P_G^{\ast}$ with covariance (\ref{Ginpd}), ${\cal R}_{\text{DF}}(P_G^{\ast})\subseteq {\cal R}_{\text{PDF}}(P_G^{\ast})\cup {\cal R}_{\text{PDF}}(P_G^{\ast\ast})\cup {\cal R}_{\text{PDF}}(P_G^{\ast\ast\ast})$ where}
\noindent
\begin{align*}
P_G^{\ast \ast}=\sum_{u\in\cal{U}}P_G^{\ast}\; \text{and}\; P_G^{\ast \ast \ast}=\sum_{v\in\cal{V}}P_G^{\ast}
\end{align*}
\begin{proof}
For a given jointly Gaussian input distribution, suppose that a point $(r_1,r_2)$ is in the region of the DF scheme but not in that of the PDF scheme. Then, we must have the minimum sum rate in each region to be one of the two middle sum rates ($S_2$ or $S_3$) so that the rate region of the DF scheme can be bigger than that of the PDF scheme. This scenario can occur only if:
\begin{itemize}
\item
$K_{12}>K_{10}$ and $K_{21}<K_{20}$, then the second sum rate $S_2$ is the minimum for both coding schemes, or
\item
$K_{12}<K_{10}$ and $K_{21}>K_{20}$, then the third sum rate $S_3$ is the minimum for both coding schemes.
\end{itemize}

Assume that the two minimum sum rates are $S_2^{\text{PDF}}$ and $S_2^{\text{DF}}$. By substituting $V=\phi$, we have $R_{\text{DF}}(P_G^{\ast})=R_{\text{DF}}(P_G^{\ast\ast})$ since then $S=U$. Now, $R_{\text{PDF}}(P_G^{\ast\ast})$ and $R_{\text{DF}}(P_G^{\ast})$ will have the same individual rates and the same $(S_1,S_2,S_4)$. The only different is in $S_3$. However, we will show that $S_2$ is the minimum for both schemes even with $V=\phi$, hence the two regions $R_{\text{PDF}}(P_G^{\ast\ast})$ and $R_{\text{DF}}(P_G^{\ast})$ are equivalent. When $V=\phi$, $S_2$ for both schemes can be expressed as
\noindent
\begin{align}\label{equpr1}
S_{2,V=\phi}^{\text{PDF}}=S_{2,V=\phi}^{\text{DF}}=&\; \alpha_1C\left(\frac{K_{10}^2\left(P_{U}+P_{10}\right)}{N}\right)+\alpha_2C\left(\frac{K_{21}^2P_{20}}{N}\right)\nonumber\\
&+\alpha_3C\left(\frac{K_{10}^2(P_{13}+c_2P_{U})+K_{20}^2(P_{23}+d_3P_{U})+2K_{10}K_{20}\sqrt{c_2d_3}P_{U}}{N}\right)
\end{align}
\noindent while $S_3$ for each scheme can be expressed as
\noindent
\begin{align}\label{equpr2}
S_{2,V=\phi}^{\text{PDF}}=&\; \alpha_1C\left(\frac{K_{12}^2\left(P_{U}+P_{10}\right)}{N}\right)+\alpha_2C\left(\frac{K_{20}^2P_{20}}{N}\right)
+\alpha_3C\left(\frac{K_{10}^2P_{13}+K_{20}^2P_{23}}{N}\right)\nonumber\\
S_{2,V=\phi}^{\text{DF}}=&\; \alpha_1C\left(\frac{K_{12}^2\left(P_{U}+P_{10}\right)}{N}\right)+\alpha_2C\left(\frac{K_{20}^2P_{20}}{N}\right)\nonumber\\
&+\alpha_3C\left(\frac{K_{10}^2(P_{13}+c_2P_{U})+K_{20}^2(P_{23}+d_3P_{U})+2K_{10}K_{20}\sqrt{c_2d_3}P_{U}}{N}\right)
\end{align}
\noindent Since we assume that $S_2$ for both original regions, which means $K_{12}>K_{10}$ and $K_{21}<K_{20}$, then from (\ref{equpr1}) and  (\ref{equpr2}), we can directly see that when $V=\phi$,  $S_{2,V=\phi}^{\text{DF}}<S_{3,V=\phi}^{\text{DF}}$.

To show that $S_{2,V=\phi}^{\text{PDF}}$ in (\ref{equpr1}) is smaller than $S_{3,V=\phi}^{\text{PDF}}$ in (\ref{equpr2}), we use the following observation: before substituting $V=\phi$, we have $K_{21}<K_{20}$ because $S_2^{\text{DF}}<S_4^{\text{DF}}$. Moreover, since $S_2^{\text{DF}}<S_1^{\text{DF}}$, we have
\begin{align}\label{equpr4}
&\alpha_1C\left(\frac{K_{10}^2\left(P_{U}+P_{10}\right)}{N}\right)+\alpha_3 C\left(\frac{K_{10}^2P_{13}+K_{20}^2P_{23}+P_U(K_{10}\sqrt{c_2}+K_{20}\sqrt{d_3})^2+P_V(K_{10}\sqrt{c_3}+K_{20}\sqrt{d_2})^2}{N}\right)
\nonumber\\
<&\alpha_1C\left(\frac{K_{12}^2\left(P_{U}+P_{10}\right)}{N}\right)+\alpha_3C\left(\frac{K_{10}^2P_{13}+K_{20}^2P_{23}}{N}\right)
\end{align}
\noindent From (\ref{equpr4}), we can see that $S_{2,V=\phi}^{\text{PDF}}$ in (\ref{equpr1}) is smaller than $S_{3,V=\phi}^{\text{PDF}}$ in (\ref{equpr2}).

Hence, after substituting $V=\phi$, $S_{2,V=\phi}^{\text{PDF}}$ is still the minimum among the other sum rates. Therefore, the the two regions $R_{\text{PDF}}(P_G^{\ast\ast})$ and $R_{\text{DF}}(P_G^{\ast})$ are equivalent and the point $(r_1,r_2)$ is in the rate region of the PDF scheme when $V=\phi$.

Similar procedure can be applied when $S_3$ is the minimum in each scheme by substituting $U=\phi$. As a result, it follows that ${\cal R}_{\text{DF}}(P_G^{\ast})\subseteq {\cal R}_{\text{PDF}}(P_G^{\ast})\cup {\cal R}_{\text{PDF}}(P_G^{\ast\ast})\cup {\cal R}_{\text{PDF}}(P_G^{\ast\ast\ast})$
\end{proof}

Finally, because of this corollary, we have $\cal{R}_{\text{DF}}\subseteq \cal{R}_{\text{PDF}}$ and since we show that
$\cal{R}_{\text{PDF}}\subseteq \cal{R}_{\text{DF}}$, we obtain the final result $\cal{R}_{\text{DF}}= \cal{R}_{\text{PDF}}$.

\bibliographystyle{IEEEtran}
\bibliography{references}

\begin{thebibliography}{10}
\providecommand{\url}[1]{#1}
\csname url@samestyle\endcsname
\providecommand{\newblock}{\relax}
\providecommand{\bibinfo}[2]{#2}
\providecommand{\BIBentrySTDinterwordspacing}{\spaceskip=0pt\relax}
\providecommand{\BIBentryALTinterwordstretchfactor}{4}
\providecommand{\BIBentryALTinterwordspacing}{\spaceskip=\fontdimen2\font plus
\BIBentryALTinterwordstretchfactor\fontdimen3\font minus
  \fontdimen4\font\relax}
\providecommand{\BIBforeignlanguage}[2]{{%
\expandafter\ifx\csname l@#1\endcsname\relax
\typeout{** WARNING: IEEEtran.bst: No hyphenation pattern has been}%
\typeout{** loaded for the language `#1'. Using the pattern for}%
\typeout{** the default language instead.}%
\else
\language=\csname l@#1\endcsname
\fi
#2}}
\providecommand{\BIBdecl}{\relax}
\BIBdecl

\bibitem{fcc2005fof}
F.~M.~J. Willems, E.~C. van~der Meulen, and J.~P.~M. Schalkwijk, ``{Achievable
  rate region for the multiple-access channel with generalized feedback},'' in
  \emph{Proc. Annu. Allerton Conf. on Communication, Control and Computing},
  1983, pp. 284--292.

\bibitem{ghasemi2007fls}
F.~Willems, ``{The discrete memoryless multiple access channel with partially
  cooperating encoders (Corresp.)},'' \emph{IEEE Trans. Inf. Theory}, vol.~29,
  no.~3, pp. 441--445, 1983.

\bibitem{haykin2005crb}
D.~Slepian and J.~Wolf, ``{A Coding theorem for multiple access channels with
  correlated sources},'' \emph{Bell Sys. Tech. Journal}, vol.~52, no.~7, pp.
  1037--1076, Sep. 1973.

\bibitem{dbout}
R.~Tandon and S.~Ulukus, ``{Dependence balance based outer bounds for Gaussian
  networks with cooperation and feedback},'' \emph{IEEE Trans. Inf. Theory},
  vol.~57, no.~7, pp. 4063--4086, Jul. 2011.

\bibitem{ray4}
A.~Hekstra and F.~Willems, ``{Dependence balance bounds for single output
  two-way channels},'' \emph{IEEE Trans. Inf. Theory}, vol.~35, no.~1, pp.
  44--53, Jan. 1989.

\bibitem{lieblein1955mos}
E.~Ekrem and S.~Ulukus, ``{Effects of cooperation on the secrecy of multiple
  access channels with generalized feedback},'' in \emph{Proc. 42nd Annu. Conf.
  on Inf. Sciences and Systems}, Mar. 2008, pp. 791--796.

\bibitem{ray1}
E.~C. van~der Meulen, ``{Three-terminal communication channels},'' \emph{Adv.
  Appl. Prob.}, vol.~3, pp. 120--154, 1971.

\bibitem{ray2}
T.~M. Cover and A.~El~Gamal, ``{Capacity theorems for the relay channel},''
  \emph{IEEE Trans. Inf. Theory}, vol.~25, pp. 572--584, Sep. 1979.

\bibitem{kolodzy2006itm}
G.~Kramer, M.~Gastpar, and P.~Gupta, ``{Cooperative strategies and capacity
  theorems for relay networks},'' \emph{IEEE Trans. Inf. Theory}, vol.~51,
  no.~9, pp. 3037--3063, Sep. 2005.

\bibitem{ray5}
Y.~Liang and V.~Veeravalli, ``{The impact of relaying on the capacity of
  broadcast channels},'' in \emph{IEEE ISIT}, Jul. 2004, p. 403.

\bibitem{ray6}
A.~Reznik, S.~Kulkarni, and S.~Verdu, ``{Broadcast-relay channel: capacity
  region bounds},'' in \emph{IEEE ISIT}, Sep. 2005, pp. 820--824.

\bibitem{cognitiveRT}
A.~Sendonaris, E.~Erkip, and B.~Aazhang, ``User cooperation diversity - part
  \uppercase{I},'' \emph{IEEE Trans. Com.}, vol.~51, no.~11, pp. 1927--1938,
  Nov. 2003.

\bibitem{sagias2004pad}
J.~N. Laneman, D.~N.~C. Tse, and G.~W. Wornell, ``Cooperative diversity in
  wireless networks: Efficient protocols and outage behavior,'' \emph{IEEE
  Trans. Inf. Theory}, vol.~50, no.~12, pp. 3062--3080, Dec. 2004.

\bibitem{simon2005dco}
S.~Vishwanath, S.~Jafar, and S.~Sandhu, ``{Half duplex relays: cooperative
  communication strategies and outer bounds},'' in \emph{IEEE Int'l Conf. on
  Wireless Net., Com. and Mobile Computing}, 2005, pp. 1455--1459.

\bibitem{nakagami1960mdg}
Y.~Peng and D.~Rajan, ``{Capacity bounds of half-duplex Gaussian cooperative
  interference channel},'' in \emph{IEEE ISIT}, 2009, pp. 2081--2085.

\bibitem{turin1972smu}
R.~Wu, V.~Prabhakaran, and P.~Viswanath, ``{Interference channels with half
  duplex source cooperation},'' in \emph{IEEE ISIT}, 2010, pp. 375--379.

\bibitem{ray7}
N.~Kim, S.~Devroye and T.~Tarokh, ``{Bi-directional half-duplex relaying
  protocols},'' \emph{journal of communications and networks}, vol.~11, no.~5,
  pp. 433--444, Oct. 2009.

\bibitem{ray8}
C.~Schnurr, S.~Stanczak, and T.~Oechtering, ``{Achievable rates for the
  restricted half-duplex two-way relay channel under a
  partial-decode-and-forward protocol},'' in \emph{IEEE ITW}, May 2008, pp.
  134--138.

\bibitem{rorth}
A.~El~Gamal and S.~Zahedi, ``{Capacity of a class of relay channels with
  orthogonal components},'' \emph{IEEE Trans. Inf. Theory}, vol.~51, no.~5, pp.
  1815--1817, May 2005.

\bibitem{hansen1977mfr}
T.~M. Cover and J.~A. Thomas, \emph{{Elements of Information Theory}},
  2nd~ed.\hskip 1em plus 0.5em minus 0.4em\relax New York:Wiley, 2006.

\bibitem{abramowitz1972hmf}
R.~G. Gallager, \emph{{Information Theory and Reliable Communication}}.\hskip
  1em plus 0.5em minus 0.4em\relax New York:Wiley, 1968.

\bibitem{haivu3}
A.~Abu Al~Haija and M.~Vu, ``{A half-duplex cooperative scheme with partial
  decode-forward relaying},'' in \emph{IEEE ISIT}, Aug. 2011.

\bibitem{haivu2}
------, ``{Throughput-optimal half-duplex cooperative scheme with partial
  decode-forward relaying},'' in \emph{IEEE ICC}, Jun. 2011.

\bibitem{hsce612}
R.~El~Gamal and Y.-H. Kim, \emph{{Lecture Notes on Network Information
  Theory}}.\hskip 1em plus 0.5em minus 0.4em\relax Available:
  http://arxiv.org/abs/1001.3404/, 2010.

\bibitem{hkintf}
H.~F. Chong, M.~Motani, H.~K. Garg, and H.~El~Gamal, ``{On the Han-Kobayashi
  region for the interference channel},'' \emph{IEEE Trans. Inf. Theory},
  vol.~54, no.~7, pp. 3188--3195, Jul. 2008.

\end{thebibliography}
\end{document}